\input vanilla.sty
\scaletype{\magstep1}
\baselineskip=14pt
\pagewidth{5.2in}

\def\p{\phantom{..}}
\def\bb{\vrule height7pt width4pt depth1pt}
 \def\NN{\text{I\kern-2pt N}}
\def\small{}
\def\comp{{\tt COMP}}
\def\x{\paper}

\centerline{\bf Computing only minimal answers in disjunctive deductive databases}

\p

\p

\centerline{\bf C A Johnson}

\centerline{School of Computing}

\centerline{University of Plymouth}

\centerline{Plymouth, PL4 8AA}

\centerline{England}

\centerline{\it  chrisj\@soc.plymouth.ac.uk}

\p

\p

\noindent
{\bf Abstract.}   A
method is presented for computing minimal answers of the form $\bigvee \Cal A$ in disjunctive deductive databases under the disjunctive stable model semantics.   Such answers are constructed by repeatedly extending partial answers.  Our method is complete (in that every minimal answer can be computed) and does not admit redundancy (in the sense that every partial answer generated can be extended to a minimal answer), whence no 
non-minimal answer is generated.      For stratified databases, the method does not (necessarily) require the computation of models of the database in their entirety.    Compilation is proposed as a tool by which problems relating to computational efficiency and the non-existence of disjunctive stable models can be overcome.   The extension of our method to other semantics is also considered.

\phantom{..}


\phantom{.ssss.}

\phantom{.ssss.}

\noindent {\bf Introduction}

\phantom{3:.....}

A propositional disjunctive deductive database [Gr86, He88, Lb92] consists of logical rules of the form
$$
A_1 \wedge A_2 \wedge \ldots \wedge A_h \wedge \lnot A_{h+1}\wedge \dots \wedge   \lnot A_{h+r} \to B_1
\vee B_2 \vee
\ldots \vee B_k
$$
where each $A_i, B_j$ is a predicate.     In such databases, answers (to queries) can themselves be disjunctive, thus an answer is a disjunction of predicates $\bigvee \Cal A$ which is logically implied (under the chosen semantics) by the database.       Of course if $\Cal A$ is non-minimal (i.e., there is some set $\Cal B\subset \Cal A$ such that $\bigvee \Cal B$ is logically implied), then $\Cal A$ is redundant, and we are therefore only interested in the computation of minimal answers.  
 
\newpage

 Minimal answer computation   is also of wider interest 
as a result of the fact that  many questions  relating to minimal answers (e.g., determining whether a given predicate belongs to some minimal answer) are $\Sigma_2^P$ - complete ([Ei93, Jo96, Mi82]), and naturally occurring problems residing at $\Sigma_n^P$ ($n\ge 2$) in the polynomial-time hierarchy are rare [St77, Wa86, Wr77].

Of course one obvious way to compute minimal answers is to compute all answers, and then eliminate non-minimal answers by subsumption.    Methods for achieving this are well known.   For example, in a positive database we can compute answers using the hyper-resolution operator:
$${\bigwedge_{i\le r} A_i \to \bigvee \Cal B;  A_i\vee \bigvee \Cal P_i \,(i\le r)}\over{\bigvee\Cal B \cup \bigcup_{i\le r}  \Cal P_i}$$
and this may be achieved by simple forward application, backward application [Ra89, Lb92], or by a combination of backward and forward [Jo98].    In each case however there is the possibility that non-minimal answers will be generated.    For example if our database consists of the rules $\{A\vee B, A\to C, B\to D, C\to D\}$, then in order to generate the minimal answer $D$, forward application of the hyper-resolution operator above {\it must} generate a non-minimal answer (i.e., $A\vee D$, $B\vee D$ or $C\vee D$).     Indeed it is fairly evident that a query answering method will always risk generating non-minimal answer if it is based upon the construction of (a structure which in effect encodes) a proof of the answer being generated:   clearly the existence of such a proof in no way precludes the possibility of a proof of some smaller disjunction using other rules within the database.

The approach presented in [Bra95] to query processing is to apply {\it unfolding} (similar to a middle-out application of the above hyper-resolution operator), which transforms a non-positive database into an equivalent set of {\it conditional facts} (i.e., rules whose bodies contain only negative atoms), and from which the set of answers can be generated using hyper-resolution.   Subsumption is again required to generate minimal answers.

An alternative approach is to generate answers by computing models of the database.  For example in [Ya94, Fe95, Fe95a], bottom-up methods for computing  minimal, perfect and disjunctive stable models are presented using the notion of a model tree.   Answers can then be derived by picking a predicate 
from each of the branches (i.e., models) in the tree [Fe95a], but again subsumption would be required to compute minimal answers.   A further disadvantage is that such an approach requires the computation of all models (of the appropriate type) in their entirety.      

An alternative approach to query processing, again using model trees,  is presented in [Ya96, Ya02], where it is observed that top-down query processing for positive databases can be achieved by the application of a model generating procedure to the dual database.    It is not clear that this approach can be extended to non-positive databases, and as noted in [Ya96], answer minimality is not addressed.

The purpose of this paper
 is to present a query answering method which, in contrast to existing approaches, generates only minimal answers.   For the most part we work under the disjunctive stable model semantics [Pr91], although the extension of our results to other semantics is considered.

Answers will be generated iteratively $A_1$,  $A_1\vee A_2$, $A_1\vee A_2\vee A_3$, ..., $A_1\vee A_2\vee \dots \vee A_n$, where, for each disjunction $A_1\vee A_2\vee \dots \vee A_r$ generated,  we guarantee that  $A_1\vee A_2\vee \dots \vee A_r$ can indeed be extended to a minimal answer by the construction  of a suitable set of {\it cyclic strong covers} [Jo99a], each of which can be viewed as a partial (disjunctive stable) model of the database.     
 Our method is complete (in the sense that every minimal answer can be 
 generated), and does not admit redundancy (in the sense that every partial answer generated can be extended to a minimal answer), whence no 
non-minimal answer is generated.    

In Section 1 we review some background and terminology, and in particular we present the conditions (in terms of disjunctive stable model membership), under which a disjunction of predicates can be extended to a minimal answer.   The computation of disjunctive stable models in their entirety is clearly undesirable, and, under certain circumstances, this can be avoided using the concept of a cyclic strong cover, which we re-introduce in Section 2.     Section 2 also presents results which demonstrate that we may  view cyclic strong covers as partial disjunctive stable models.

  In Section 3 we focus on stratified databases.  For such databases, disjunctive stable models coincide with perfect models, whence cyclic strong covers may be viewed as partial perfect models.   More importantly, cyclic strong covers can always be extended to full perfect models, and hence provide a characterisation of  reasoning under the perfect model semantics.  This then allows a characterisation of partial minimal answers in terms of cyclic strong covers.  Using this, a method for constructing minimal answers is derived in terms of repeated extension of cyclic strong covers, and moreover this can be achieved without necessarily constructing perfect models in their entirety. 

In Section 4 we consider unstratified databases.  For such databases, total cyclic strong covers coincide with disjunctive stable models, but  the possible non-existence of disjunctive stable models means that (non-total) cyclic strong covers cannot always be extended to full disjunctive stable models.     This suggests that partial minimal answers can only be witnessed as such using total cyclic strong covers, and (hence) that a direct extension of the methods of Section 3 to unstratified databases  requires the (undesirable) computation of disjunctive stable models in their entirety.   

In Section 5 we show that this problem can be addressed by  partitioning the database into extensional and intensional components,  and  pre-processing  ({\it compiling}) the construction of cyclic strong covers within the intension.    This  greatly simplifies, and hence reduces the cost of, the run-time computation (which then takes place within the extensional database), this saving being of particular importance given the computational complexity of the problem.   An important property of our compilation is that it does not need to be repeated following updates to the extensional database.     

In Section 6 we question whether the techniques of the present paper might, from the viewpoint of computational efficiency, be usefully combined with other techniques that are themselves known to generate non-minimal answers.  We briefly  consider the adaptation of our methods to other semantics  in Section 7, and in Section 8 consider the issues that arise when lifting our methods to the first order level.   Our conclusions and suggestions for further research are presented in Section 9.  An extended worked example is presented in Appendix A, and an alternative form of pre-processing is presented briefly in Appendix B.

\p

\p

\noindent {\bf \S 1. \,\,  Minimal answers}

\p

In this section we review some background and terminology, and in particular we state the conditions (in terms of disjunctive stable model membership), under which a disjunction  of predicates can be extended to a minimal answer.   

\p

\noindent {\bf 1.1 Notation.}   Throughout we assume that $\Cal L$ is a finite propositional language (i.e., a finite set of predicates).     A literal is a predicate (a positive literal) or its negation (a negative literal), and we will use $\Cal A, \Cal B, \Cal C, \dots, \Cal P, \Cal Q, \dots$ to denote arbitrary sets of literals.     $\Cal Q^-=\{P\in \Cal L \vert \lnot P \in \Cal Q\}$, $\Cal Q^+=\{P\in \Cal L \vert P \in \Cal Q\}$ and $\overline{\Cal Q}=\{\lnot K \vert K \in \Cal Q\}$.   $\Cal Q$ is {\it total} iff $\Cal Q^-\cup \Cal 
Q^+=\Cal L$, and {\it consistent} iff $\Cal Q^-\cap \Cal Q^+=\emptyset$.

Throughout $T$ will denote a  disjunctive deductive database in $\Cal L$ consisting of rules $C$ of the form 
$
A_1 \wedge A_2 \wedge \ldots \wedge A_h \wedge \lnot A_{h+1}\wedge \dots \wedge   \lnot A_{h+r} \to B_1
\vee B_2 \vee
\ldots \vee B_k
$, where each $A_i, B_j$ is a predicate and $k>0$.     We may assume without loss of generality that if $r>0$, then $h>0$.

antec($C) = \{A_1, A_2, \ldots, A_h \}$ denotes the set of {\it antecedents} of $C$, $\Cal N(C) = \{A_{h+1},$ $ A_{h+2}, \dots, A_{h+r}\}$, and conseq($C)=\{B_1, B_2, \dots, B_k\}$ denotes the {\it consequent} of $C$.   $T$ is said to be {\it positive} iff $\Cal N(C) = \emptyset$ for each $C\in T$.

\p

\noindent {\bf 1.2 Definition.}   A set $M\subseteq \Cal L$ is a {\it model} of $C$ (written $M\models C$) iff antec($C)\subseteq M$  and $M\cap \Cal N(C)=\emptyset$ implies that conseq($C)\cap M \ne \emptyset$.   $M$ is a model of $T$ (written $M\models T$) iff $M\models C$ for each $C\in T$. 

\p

A total consistent  set of literals identifies a truth value for each predicate, which in turn allows us to determine truth values for rules.   Given a consistent set of literals $\Cal C$, we let $T/\Cal C$ denote those rules in $T$ whose predicates are given a truth value by $\Cal C$.  

\p

 \noindent {\bf 1.3 Definition.}     If $\Cal C$ is a consistent  set of literals,  let  $T/\Cal C = \{C\in T \vert \,\,  \text{conseq}(C)\cup \text{antec}(C)\cup \Cal N(C)\subseteq \Cal C^+\cup \Cal C^-\}$.

\p

\noindent {\bf 1.4 Definitions.}   
\item{\bf (a)}   If $C$ is a rule, let $pos(C) = \bigwedge \text{antec}(C) \, \to \, \bigvee \text{conseq}(C)$, i.e., $pos(C)$ is formed from $C$ by removing the negative literals from the body of $C$.
\item{\bf (b)}  If $N\subseteq \Cal L$, then the {\it Gelfond-Lifschitz} transformation [Ge88] is given by $T\vert_g N = \{pos(C) \vert \, C\in T,  \Cal N(C) \cap N = \emptyset\}$.    Notice that $T\vert_g N$ is positive, and can therefore be interpreted straightforwardly using the minimal model semantics.
\item{\bf (c)}  If $M\subseteq \Cal L$, then $M$ is a {\it disjunctive stable model} of $T$ [Pr91] iff $M$ is a minimal model of $T\vert_g M$.     
\item{\bf (d)}  If $\Phi$ is a formula in $\Cal L$, then we write $T\models \Phi$ iff $\Phi$ is true in every disjunctive stable model of $T$. 
\item{\bf (e)} If $\Cal A \subseteq \Cal L$, then 
$\bigvee\Cal A$ is a {\it minimal answer} in $T$  iff $T\models \bigvee \Cal A$ and there is no proper subset $\Cal B\subset \Cal A$ such that $T\models \bigvee\Cal B$.   

\p

Disjunctive stable models are a straightforward generalisation of stable models for non-disjunctive databases  [Ge88].    Note that every disjunctive stable model of $T$ is a minimal model of $T$, and that for positive databases the converse holds.

Given a minimal answer $\bigvee\Cal A$, we will for the sake of brevity also refer to  the set $\Cal A$ as a minimal answer.   If $\bigvee \Cal A$ is a minimal answer and $A\in \Cal A$, then there must be some disjunctive stable model $M$ of $T$ such that $M\cap \Cal A = \{A\}$ (for otherwise $T\models \bigvee (\Cal A - \{A\})$).    Conversely if $A$ belongs to a disjunctive stable model $M$, then by the minimality of $M$ we must have that $T\models A \vee \bigvee (\Cal L - M)$, whence if $\Cal A\subseteq \{A\}\cup (\Cal L - M)$ is a minimal answer, then $\emptyset \ne M\cap \Cal A \subseteq \{A\}$, and hence $A\in \Cal A$.    This then yields the following which is a direct analogue of a result originally presented (for the minimal model semantics) by  Minker. 

\p

\noindent {\bf 1.5 Theorem}  [Mi82].    A predicate $A$ belongs to some minimal answer (in $T$) iff $A$ belongs to some disjunctive stable model of $T$.

\p

If we are to try to construct minimal answers as a sequence $A_1$, $A_1\vee A_2$, ..., then Theorem 1.5 dictates that each such $A_i$ needs to belong to some disjunctive stable model $M_i$.     The following theorem indicates that we also need to consider the relationship between the models $M_i$.

\p

\noindent {\bf 1.6 Theorem.}     A set of predicates $\{A_i \vert i \le r\} $ is contained in a minimal answer iff for each $i\le r$ we may find a disjunctive stable model $M_i$ of $T$ such that 
\item{\bf (a)}   $M_i\cap \{A_j \vert j \le r\}=\{A_i\}$, and 
\item{\bf (b)}  $T\models \bigvee_{i\le r} A_i \vee \bigvee \bigcap_{i\le r} (\Cal L-M_i)$.

\p

\noindent {\bf Proof} ($\to$).    Suppose that $A_1\vee A_2\vee\dots \vee A_{r+s}$ is a minimal answer, then for each $i\le r$ we may find  a disjunctive stable  model $M_i$ such that $M_i\cap\{A_j \vert j\le r+s\}=\{A_i\}$.    If $M$ is a disjunctive stable model of $T$ with $M\not\models A_1\vee A_2\vee\dots \vee A_r$, then there is some $j>r$ such that $A_j\in M$, where $A_j\in \bigcap_{i\le r} (\Cal L-M_i)$.

\noindent $(\leftarrow)$.    Let $\Cal A\subseteq \{A_i \vert i \le r\} \cup \bigcap_{i\le r} (\Cal L-M_i)$ be a minimal answer, then for each $i\le r$, $\emptyset \ne M_i\cap \Cal A \subseteq \{A_i\}$, whence  $A_i\in \Cal A$.   \bb

\p

\p

\noindent {\bf \S 2.   \,\,  Cyclic strong covers}

\p

Theorem 1.6 characterises partial minimal answers in terms of disjunctive stable models.    Ideally however we would prefer not to have to construct models of the database in their entirety.  With this in mind, this section re-introduces cyclic strong covers [Jo99a] which we will show function (to some extent) as partial disjunctive stable models, and which allow an appropriate re-formulation of Theorem 1.6.   
Strong covers provide the notion of model-hood, and cyclicness  captures minimality. 

\p

\noindent {\bf 2.1 Definition} [Jo99a].  Let $\Cal
Q$ be a consistent set of literals in $\Cal L$.   A {\it strong cover} of $\Cal Q$ (in $T$)
is a consistent set of literals $\Cal C\supseteq \Cal Q$ such that for each $C \in T$
$$\text{conseq}(C)\subseteq \Cal C \, \Longrightarrow \,
(\text{antec}(C) \cup \overline {\Cal N(C)}) \cap \Cal C \ne
\emptyset.$$

\p

Notice that the above definition suggests a top-down construction of strong covers.   Clearly if $\Cal C$ is a strong cover, then $\Cal L - \Cal C^+\models T$.   Conversely if $M\subseteq \Cal L$ then $M$ is a model of $T$ iff $\overline {M}\cup (\Cal L - M)$ is a strong cover in $T$.    

In [Jo96] we introduced the notion of a cyclic tree\footnote{Throughout this paper we will use what is referred to in [Jo96] as {\it unfactored} cyclic trees.}  which facilitates reasoning about minimal, perfect and disjunctive stable models.      We first present an example to motivate the following definition of such trees, and then detail  their essential properties.     Further examples, motivation, and details of the top-down construction of such trees are discussed at length in [Jo96, Jo98, Jo98a, Jo99, Jo99a].

\p

\def\hh{\hskip-2pt}
\noindent {\bf 2.2  Example.}  Suppose that $T$ consists of the
following rules:
$$\vbox{\settabs 9 \columns
\+1.  $Q_2 \wedge Q_3\wedge \lnot R_1 \to Q_1 \vee Q_5$&&&\,\,\,\,\,\,\,\,\,2. 
$Q_1
\wedge \lnot R_2 \to Q_2 $&&&3.  $S_2\wedge \lnot R_3  \to
Q_3$&\cr
\+4.  $S_3 \to Q_1\vee Q_2\vee Q_6 $&&&\,\,\,\,\,\,\,\,\,5.  $S_2\vee R_5$&&&6. 
$S_1
\to  Q_3 \vee Q_2$ \cr
\+7.  $S_3\vee R_7$&&&\,\,\,\,\,\,\,\,\,8.  $Q_5\wedge R_1 \to Q_2$&&&\cr
}$$
and we wish to determine whether $Q_1$ lies in some disjunctive stable model 
of $T$.        

Suppose that $Q_1$ lies in the disjunctive stable model  $M$, then by the minimality of $M$ 
we may find
a rule $C\in T$ such that $\Cal N(C)\cap M =\emptyset$ (i.e., $pos(C)\in T\vert_g M$) and $M-\{Q_1\}\not\models pos(C)$.  There are only two
possibilities for $C$, namely rules 1 and 4.     Suppose we 
guess 
that $C$ is rule 1,
then
$\{Q_1, Q_2, Q_3\} \subseteq M \subseteq \Cal L - \{Q_5, R_1\}$.
We will represent this application of rule 1 using the ``rule node'' $rn_1$ in the 
tree $\Cal T_1$ (Figure 2.2(i)).    (Only $Q_2$ and $Q_3$ (i.e. the
antecedants) are depicted, since we wish to examine these predicates
further.)

Suppose now that we apply the same argument to $Q_2$.   If $C$
is a rule in $T$ such that $\Cal N(C)\cap
M=\emptyset$ and $M-\{Q_2\}  \not\models pos(C)$, then
(since $\{Q_1, Q_2, Q_3\} \subseteq M \subseteq \Cal L - \{Q_5, R_1\}$)
$C$ {\it must} be rule 2, thus yielding
the tree $\Cal T_2$ (Figure 2.2(i)), and the further constraint that 
$\{Q_1, Q_2, Q_3\} \subseteq M \subseteq \Cal L - \{Q_5, R_1, R_2\}$.

The left hand branch of $\Cal T_2$ forms a ``cycle''.  There is
no point working with $Q_1$ or $Q_2$ alone (since we have already
done so), thus we look for a rule $C\in T$ such that  $\Cal N(C)\cap
 M=\emptyset$ and $M-\{Q_1, Q_2\} \not \models pos(C)$, the only candidate being rule
4.  Rule 7 then
terminates the branch (since rule 7 has no antecedants), thus
yielding $\Cal
T_3$ (Figure 2.2(ii)), and the new constraint 
$\{Q_1, Q_2, Q_3, S_3\} \subseteq M \subseteq \Cal L - \{Q_5, R_1, R_2, 
Q_6, R_7\}$.

\baselineskip=20pt
$$\vbox{\settabs 9 \columns
\+&&\,\,$Q_1$&&&&\,$Q_1$&&\cr
\+&&\hh\,\, $rn_1$&&&&\hh $rn_1$&\cr
\+&$Q_2$&&$Q_3$&&\,$Q_2$&&$Q_3$&\cr
\+&&$\Cal T_1$&&&\hh\, $rn_2$&&&&\cr
\+&&&&&\,\,$Q_1$&&&\cr
\+&&&&&&$\Cal T_2$&&\cr
}$$
\baselineskip=14pt
$$\text{\bf Figure 2.2(i).}$$

\baselineskip=20pt
$$\vbox{\settabs 9 \columns
\+&&\,\,\,\,$Q_1$&&&&\,\,$Q_1$&&\cr
\+&&\hh\,\, $rn_1$&&&&\hh\,\, $rn_1$&\cr
\+&$Q_2$&&\,\,\,\,$Q_3$&&$Q_2$&&\,$Q_3$&\cr
\+&\hh\, $rn_2$&&&&\hh $rn_2$&&\hh $rn_3$\cr
\+&\,$Q_1$&&&&\,$Q_1$&&\,$S_2$\cr
\+&\hh\, $rn_4$&&&&\hh $rn_4$&&\hh $rn_5$\cr
\+&\,$S_3$&&&&\,$S_3$&&&&\cr
\+&\hh\, $rn_7$&&&&\hh $rn_7$&&&\cr
\+&&$\Cal T_3$&&&&$\Cal T_4$&&\cr
}$$
\baselineskip=14pt
$$\text{\bf Figure 2.2(ii).}$$

We thus move on to examine $Q_3$: if $M-\{Q_3\} \not
\models pos(C)$, then (given the existing constraints on $M$) $C$ must be rule
3.   Rule 5 can then be used to handle
$S_2$, and again this 
terminates the branch, yielding $\Cal T_4$ (Figure 2.2(ii)),
and the constraint
$\{Q_1, Q_2, Q_3, S_3, S_2\} \subseteq M \subseteq \Cal L -
\{Q_5, R_1, R_2, Q_6, R_7, R_3, R_5\}$.

Thus we are then left with the problem of finding 
a disjunctive stable model $M$ satisfying the above
constraint.  This appears harder than our initial problem (i.e.,
of finding such 
a model containing $Q_1$).  Note however that if $M$ is
{\it any} (disjunctive stable) model of $T$ that is
disjoint
from $\{Q_5,  R_1, R_2, Q_6, R_7, R_5, R_3\}$, then (using the
rules
in the tree) we may infer that $\{Q_1, Q_2, Q_3, S_3, S_2\}
\subseteq M$.  In particular, if we can show that
$T\not\models \bigvee\{Q_5,  R_1, R_2, Q_6, R_7, R_5, R_3\}$, then
we may
infer the existence of a disjunctive stable  model
containing $Q_1$.

\phantom{..}

Notice in the tree construction above, that it is the intention that each predicate lies in the intended model $M$.  Each ``rule node''
is
labelled with a rule $C\in T$ such that $\Cal N(C)\cap
 M=\emptyset$, and if $rn_C$ has
parent $n$, then
there
is a subset $\Cal P$ of the predicates above $n$ such that 
$M-\Cal P\not\models pos(C)$.   Thus, antec($C)\subseteq M - \Cal P$, 
and $\emptyset = \text{conseq}(C)\cap (M-\Cal P) = M \cap
(\text{conseq}(C)-\Cal P)$.   Since $M\models pos(C)$ we must have that conseq($C)\cap \Cal P
\ne \emptyset$.    

The following definition captures these features, and also
provides a precise definition of the subset $\Cal P = $ CYC($n)$ (the ``cycle above $n$'') 
that we wish to work with, this choice being motivated by the properties (Theorem 2.4) of the resulting trees.     [Jo96] compares this choice of $\Cal P$ with a number of alternatives.

\p

\noindent {\bf 2.3 Definition} [Jo96].  If $P$ is a predicate, then a {\it cyclic tree} for $P$ in $T$ contains rule nodes and
predicate nodes,  satisfying conditions (i) - (v) below.

\item{\bf (i)}  Each rule node $rn$ is labelled with a rule
$C\in T$ (written $rn_C$).   Each predicate node $n$ is
labelled with a predicate $R\in \Cal
L$ (and we write $lab(n)=R$).    We set $Pred(\Cal T) =\{lab(n) \mid n \text{
is a predicate node in } \Cal T\}$.

\item{\bf (ii)}  The root node (at the top of the
tree)
is a predicate node labelled with $P$.   
 
\item{\bf (iii)}  If $rn_C$ is a rule node,
then for each predicate $R\in$\,antec($C$), $rn_C$ has a child
node
labelled with $R$. $rn_C$ has no other
child nodes.   If $rn_C$ has parent $n$, then conseq($C)\cap \text{CYC}(n)\ne\emptyset$ and antec($C)\cap \text{CYC}(n)=\emptyset$, where $\text{CYC}(n) = \{lab(m) \mid m\ge n,  \exists n' (n' \ge m,
lab(n') = lab(n))\}.$  \newline \indent Let  
$\Cal O(rn_C)
= \text{conseq}(C) - \text{CYC}(n)$.

\item{\bf (iv)}  If $n$ is a predicate node, then $n$ is not  a leaf node, and its child is a rule node satisfying (iii) above.   If $n$ is not the  root, then its parent node
is a rule
node $rn_D$ (with $lab(n)\in \text{antec}(D)$). 
\item{\bf (v)}  $\Cal O(\Cal T) =
\bigcup
\{\Cal O(rn_C) \mid rn_C $ occurs in $\Cal T\}$ and 
$\Cal N(\Cal T) = \bigcup \{\Cal N(C) \mid rn_C $ occurs in $\Cal
T\}$ are both disjoint from $Pred(\Cal T)$.
\newline \indent 
Let $\Cal S(\Cal T)=\overline{Pred(\Cal T)}\cup \Cal O(\Cal T) \cup \Cal N(\Cal T)$.

\p

Notice that the definition of cyclic trees is inherently top-down.   
It is easy to show that the size of cyclic  trees is limited [Jo96, Section 5], in that every branch through a cyclic tree has length at most $\vert \Cal L\vert * (\vert \Cal L \vert + 1)/2$.   
Clearly  if $\Cal T$ is a cyclic tree in $T$, then $\Cal T$ is also a cyclic tree in any superset of $T$.     Notice also that if $C$ is a rule labelling some rule node in $\Cal T$, then each predicate  appearing in $C$ appears in $Pred(\Cal T) \cup \Cal O(\Cal T) \cup \Cal N(\Cal T)$.   (In the terminology of Definition 1.3,  $C\in T/\Cal S(\Cal T)$.)

The following theorem captures the essential properties of cyclic trees.

\p

\noindent {\bf 2.4  Theorem}  [Jo96, Jo99a].

\item{\bf (a)}    If the predicate  $P$ belongs to some disjunctive stable model $M$, then we may find a cyclic tree $\Cal T$ for $P$ in $T$ such that $Pred(\Cal T) \subseteq M \subseteq \Cal L - (\Cal O(\Cal T) \cup \Cal N(\Cal T))$.  

\item{\bf (b)}  If $\Cal T$ is a cyclic tree in $T$ and $M\models T\vert_g (\Cal L - \Cal N(\Cal T))\wedge \lnot \bigvee \Cal O(\Cal T)$, then $Pred(\Cal T) \subseteq M$.
\item{\bf (c)}     If $\Cal T$ is a cyclic tree in $T$ and $M\models T \wedge \lnot \bigvee (\Cal O(\Cal T) \cup \Cal N(\Cal T))$, then $Pred(\Cal T) \subseteq M$.

\p

Thus if $M$ is a disjunctive stable model of $T$, then we can find a set $\{\Cal T_i \vert i \le m\}$ of cyclic trees in $T$ such that $\bigcup_{i\le m} Pred(\Cal T_i) = M$ and $M \subseteq \Cal L - \bigcup_{i\le m} (\Cal O(\Cal T_i) \cup \Cal
N(\Cal T_i))$, i.e., $\Cal L - M \supseteq \bigcup_{i\le m} (\Cal O(\Cal T_i) \cup \Cal
N(\Cal T_i))$.  $\overline{M}\cup (\Cal L - M)$ thus has the property identified in the following  definition.
   
\p

\noindent {\bf 2.5 Definition} [Jo99].  Let
$\Cal
C$ be a consistent set of literals, then $\Cal C$ is said to
be 
{\it  cyclic} (in $T$) iff 
there is a set of cyclic trees $\{\Cal T_i \mid
i\le m\}$ in $T$ such that 
\item{\bf (i)}  $\Cal C^- = \bigcup_{i\le m} Pred(\Cal T_i)$, and 
\item{\bf (ii)}  $\Cal C^+ \supseteq \bigcup_{i\le m} (\Cal O(\Cal T_i) \cup \Cal
N(\Cal T_i))$.

\p 

It is in fact possible to allow some pruning of the cyclic trees within the definition/ construction of a cyclic set (see [Jo99a, Theorem 4.1]) although we will not consider this issue in this  paper. Again note that if $\Cal C$ is cyclic in $T$, then $\Cal C$ is also cyclic in any superset of $T$. 

Where there is no ambiguity (or required emphasis), we will, for the sake of brevity, refer to a cyclic set/strong cover in $T$ as simply cyclic/a strong cover. 

Following on from Theorem 2.4 we see for example that if $\Cal C$ is cyclic, then any model of $T\vert_g (\Cal L - \Cal C^+)\wedge \lnot \bigvee \Cal C^+$ will contain $\Cal C^-$.     In particular, if $\Cal C$ is a total cyclic strong cover with $M=\Cal C^- = \Cal L - \Cal C^+$, then $M\models T$ (whence $M\models T\vert_g M$) and if $M^*\subseteq M$ with $M^*\models T\vert_g M$, then $M^*\models T\vert_g (\Cal L - \Cal C^+)\wedge \lnot \bigvee \Cal C^+$, whence $\Cal C^-=M\subseteq M^*$.   Thus $M=M^*$ and $M$ is a disjunctive stable model of $T$.    Conversely, as indicated above, if $M$ is a disjunctive stable model, then $\overline{M}\cup (\Cal L - M)$ is a total cyclic strong cover, whence we have part (a) of the following theorem.

\p

\noindent {\bf 2.6 Theorem}  [Jo99a].  
\item{\bf (a)}  If $M\subseteq \Cal L$, then $M$
is a disjunctive stable model of $T$ iff $\overline{M} \cup (\Cal
L - M)$
is a cyclic
strong cover in $T$.
\item{\bf (b)}    If $\Cal Q$ is a set of literals, then $T\models \bigvee \Cal Q$ iff there is no total cyclic strong cover of $\Cal Q$ in $T$.
\item{\bf (c)}   A consistent set of literals $\Cal C$ is cyclic in $T$ iff  for each $P\in \Cal C^-$ there is a cyclic tree
$\Cal T$ for
$P$ in $T$ such that
$\Cal S(\Cal T)\subseteq \Cal C$.

\p

Note that if $\Cal C$ is cyclic in $T$ and $\Cal T$ is a 
cyclic tree  with $\Cal S(\Cal T) \subseteq \Cal C$, then for each 
rule $C$ labelling a rule node in $\Cal T$ we have that $C\in T/\Cal S(\Cal T) \subseteq T/\Cal C$ and 
$\Cal C^-\models C$ (since conseq($C)\cap Pred(\Cal T) \ne\emptyset$).  Thus 
$\Cal C$ is also cyclic in $\{C\in T/\Cal C \,\,\vert\,\, \Cal C^-\models C\}$.
We can thus  characterise cyclicness in terms of disjunctive stable models. 

\p

\noindent {\bf 2.7 Theorem.}  
\item{\bf (a)}  A consistent set of literals $\Cal C$ is cyclic in $T$ iff $\Cal C^-$ is a disjunctive stable model of $\{C\in T/\Cal C \,\,\vert\,\, \Cal C^-\models C\}$.
\item{\bf (b)}  If $\Cal C$ is a strong cover in $T$, then $\Cal C$ is cyclic in $T$ iff $\Cal C^-$  is a disjunctive stable model of $T/ \Cal C$.

\p

\noindent {\bf Proof (a).}  ($\to$).   By the above remark,   $\Cal C$ is a total cyclic   strong cover in $\{C\in T/\Cal C \,\,\vert\,\, \Cal C^-\models C\}$.

\noindent $(\leftarrow$).   Every predicate in $\{C\in T/\Cal C \,\,\vert\,\, \Cal C^-\models C\}$ appears in $\Cal C^+\cup \Cal C^-$.  Thus by Theorem 2.6(a),  $\Cal C = \overline {\Cal C^-}\cup ((\Cal C^+\cup \Cal C^-)-\Cal C^-)$ is cyclic in $\{C\in T/\Cal C \,\,\vert\,\, \Cal C^-\models C\}$, whence cyclic in $T$.

\noindent {\bf (b).}   If $\Cal C$ is a strong cover in $T$, then $\Cal C^-\models T/\Cal C$, whence the result follows trivially from part (a).   \bb 

\p

Theorems 2.6(a) and 2.7 indicate that cyclic strong covers can in some sense be regarded as partial disjunctive stable models.   Because of the possible non-existence of disjunctive stable models, it is of course not the case that every cyclic strong cover can be extended to a disjunctive stable model.   By Theorem 2.6(b), if $\Cal C$ is a cyclic strong cover, then $\Cal C$ may be extended to a total cyclic strong cover (i.e., a disjunctive stable model) iff $T\not\models \bigvee \Cal C$, and by Theorem 2.4(c) this is the case  iff $T\not\models \bigvee \Cal C^+$.    We can also  show (Theorem 2.9 below) that the disjunctive stable models extending a cyclic strong cover are precisely the disjunctive stable models of some reduced database, and thus  that an inability to extend cyclic strong covers is simply the phenomena of the non-existence of such models in a different guise.  Definition 2.8 below defines our reduced database.

\p

\noindent {\bf 2.8 Definition.}  If $\Cal D$ is a strong cover in $T$ and  $C$ is rule in $T$, then  let $C_{\Cal D}$ be formed from $C$ by replacing each predicate in $\Cal D^-$ by {\tt TRUE}, and each predicate in $\Cal D^+$ by {\tt FALSE}.

Note that if conseq($C)\cap \Cal D^-\ne\emptyset$, then $C_{\Cal D}$ reduces to {\tt TRUE}.  Also, if conseq($C)\subseteq \Cal D^+$ then conseq($C_{\Cal D})$ reduces to {\tt FALSE}, but since $\Cal D$ is a strong cover there must be some literal in the body of $C$ which evaluates to {\tt FALSE}, whence $C_{\Cal D}$ as a whole reduces to {\tt TRUE}.   

In general $C_{\Cal D}$ reduces to {\tt TRUE} ($C_{\Cal D}\equiv {\tt TRUE}$)
iff conseq$(C)\cap \Cal D^-\ne\emptyset$    or    $\text{antec}(C)\cap \Cal D^+\ne\emptyset$ or $ \Cal N(C)\cap \Cal D^-\ne\emptyset.$  

Let $T_{\Cal D} = \{C_{\Cal D} : C\in T, C_{\Cal D} \not\equiv {\tt TRUE}\}$. Clearly a rule $C^*$ is in $T_{\Cal D}$ iff there is a rule $C\in T$ such that  
\item{\bf (i)}   conseq($C)\cap \Cal D^- = \emptyset$ and conseq($C^*) = \text{conseq}(C) - \Cal D^+$,  
\item{\bf (ii)}  antec($C)\cap \Cal D^+= \emptyset$ and antec($C^*)=\text{antec}(C)-\Cal D^-$, and 
\item{\bf (iii)}   $\Cal N(C)\cap \Cal D^-=\emptyset$ and $\Cal N(C^*) = \Cal N(C) - \Cal D^+$

\noindent and that under these conditions we have that $C^*=C_{\Cal D}$. 
Notice that only predicates from $\Cal L - (\Cal D^+\cup \Cal D^-)$  appear in $T_{\Cal D}$.   

\p

\noindent {\bf 2.9  Theorem.}    Suppose that $\Cal D$ is a cyclic strong cover in $T$ and $\Cal D^-\subseteq M \subseteq \Cal L - \Cal D^+$.      Then $ M$ is a disjunctive stable model of $T$ iff $M - \Cal D^-$ is a disjunctive stable model of $T_{\Cal D}$.

\p

\def\paper{The proof is not especially instructive, and we thus delegate it to an accompanying technical report [Jo03a].}

\x

Since cyclic strong covers are themselves disjunctive stable models of some sub-database, the theorem above also allows us to characterise when one cyclic strong cover extends another.

\p

\noindent {\bf 2.10  Corollary.}  Suppose that $\Cal D$ is a cyclic strong cover in $T$ and $\Cal G$ is a consistent set of literals with $\Cal G\supseteq \Cal D$.   

\noindent {\bf (a)}   $\Cal G$ is a strong cover in $T$ iff  $\Cal G-\Cal D$ is strong cover in $T_{\Cal D}$.

\noindent {\bf (b)}    $\Cal G$ is a cyclic strong cover in $T$ iff  $\Cal G-\Cal D$ is cyclic strong cover in $T_{\Cal D}$.

\p

\def\paper{The proof is again delegated to the accompanying technical report [Jo03a].}

\x

As mentioned earlier, the (top-down) construction of cyclic trees is detailed in [Jo96].    The (top-down) construction of cyclic strong covers is detailed in [Jo98, Jo99, Jo99a], using the operators detailed below in Definition 2.11, and representing the cyclic strong covers generated as the branches of an {\it extended deduction tree}.     A  simple algorithm for the construction of cyclic strong covers can then be presented, based upon the use of two stacks, and operating in space that is quadratic in $\vert \Cal L\vert$ (the size of the language) [Jo98].   Appendix A contains an example of a similar such tree construction.

\p

\noindent {\bf 2.11  Definition.}      Suppose that $\Cal Q$ is cyclic, then a strong cover $\Cal C$ of $\Cal Q$ is said to be a {\it constructible extension} of $\Cal Q$ iff we can find a sequence $\Cal Q = \Cal Q_0\subseteq \Cal Q_1\subseteq \dots \subseteq \Cal Q_r = \Cal C$ ($r\ge 0$) such that for each $i\le r$ there is a rule $C_i\in T$ such that conseq($C_i)\subseteq \Cal Q_{i-1}$, $\Cal Q_{i-1}\cap (\text{antec}(C_i)\cup \overline{\Cal N(C_i)}) =  \emptyset$ and  either 
\item{\bf (i)}   $\Cal Q_i=\Cal Q_{i-1}\cup \{A_i\}$ for some  $A_i\in \text{antec}(C_i)$, or 
\item{\bf (ii)}   $\Cal Q_i=\Cal Q_{i-1}\cup \Cal S(\Cal T)$, where $\Cal T$ is a cyclic tree  for some  $A_i\in \Cal N(C_i)$ in $T$ (such that $\Cal Q_{i-1}\cup \Cal S(\Cal T)$ is consistent).

If $\Cal Q = \Cal R\cup \{\lnot A\}$, where $\Cal R$ is cyclic  and $A\not\in \Cal R^-\cup \Cal R^+$, then a constructible extension of $\Cal Q$ is formed by taking a constructible extension of $\Cal Q'=\Cal R \cup \Cal S(\Cal T)$, where $\Cal T$ is a cyclic tree for $A$ in $T$ such that $\Cal Q'$ is consistent.

\p

It will not prove necessary to define constructible extensions for sets more complex than those above.    Theorem 2.6(c) implies that constructible extensions are themselves cyclic strong covers, and also yields the following corollary.

\p

\noindent {\bf 2.12 Corollary.}   Suppose that $\Cal D$ is a cyclic strong cover of $\Cal Q$, then we may find a constructible extension $\Cal C$ of $\Cal Q$ such that $\Cal C \subseteq \Cal D$.

\p

The construction of cyclic strong covers can (by Theorem 2.6(a)) be amended to compute disjunctive stable models by  implicitly adding the denial rules $\{P\wedge \lnot P \to {\tt FALSE} \vert P\in \Cal L\}$ to $T$ (thus forcing all strong covers of $\{{\tt FALSE}\}$ to be total).    This approach  to the computation of disjunctive stable models differs from that presented in [Fe95], in that it is top-down, and moreover does not require the database to be transformed within an extended language.   Such computation of disjunctive stable models can be enhanced by the application of the results presented in Theorem 2.9 and Corollary 2.10.   

\p

\p

\noindent {\bf \S 3. \,\, Stratified databases}

\p

For stratified databases, cyclic strong covers provide a characterisation of  reasoning under the perfect model semantics [Jo99].  We will show that they can be  viewed as partial perfect models, and we then use cyclic strong covers to define the notion of a {\it cyclic state} which provides a characterisation of partial minimal answers.   Our method for constructing minimal answers is then presented in terms of an extension mechanism for cyclic states.  As a result of the fact that cyclic strong covers can always be extended to full perfect models, this can be achieved without necessarily computing perfect models in their entirety.

\p

\noindent {\bf 3.1  Definitions.}
Throughout Section 3 we assume that $T$ is {\it stratified} [Pr88], by which we mean that there is a level function $\ell : \Cal L \to \NN$ such that for each rule 
$
A_1 \wedge A_2 \wedge \ldots \wedge A_h \wedge \lnot A_{h+1}\wedge \dots \wedge   \lnot A_{h+r} \to B_1
\vee B_2 \vee
\ldots \vee B_k
$ in $T$: (i)  $\ell(B_j) = \ell(B_1)$ for each $j\le k$,  (ii)  $\ell(A_i) \le \ell(B_1)$ for each $i\le h$, and (iii)  $\ell(A_{h+i})<\ell(B_1)$ for each $i\le r$.  We define $\ell(C)=\ell(B_1)$.

If $T$ is stratified, then a model $M$ of $T$ is {\it perfect} [Pr88] iff for each $\alpha$, there is no set $M'\subset \{P\in M \vert \ell(P)=\alpha\}$ such that 
$\{P\in M \vert \ell(M)<\alpha\}\cup M'\models \{C \in T \vert \ell(C) = \alpha\}$.    It is well known [Pr91] that for stratified databases, perfect and disjunctive stable models coincide.

Theorem 1.6 dictates the requirements on a disjunction if it is to be extended to a minimal answer.   As mentioned earlier however, we certainly do not wish to (have to) compute perfect models in their entirety, and in this respect the following result is key.

\p

\noindent {\bf 3.2 Theorem}  [Jo99].  Let $\Cal Q$ be a set of
literals, then $T\models \bigvee \Cal Q$ iff  $\Cal Q$ has no cyclic strong cover in
$T$.

\p

Theorem 2.6(a) dictated that perfect models coincide with total cyclic strong covers.  If $\Cal C$ is a cyclic strong cover then (applying Theorem 3.2 with $\Cal Q = \Cal C$) we have that  $T\not\models \bigvee \Cal C$, whence there is a perfect model $M$ of $T$ such that $M\not\models \bigvee \Cal C$,  i.e., $\Cal C\subseteq \overline{M}\cup (\Cal L - M)$.     Thus (as in Section 1) cyclic strong covers may be regarded as partial perfect models, but crucially in this case they may always being extended to a full perfect model.

We now seek to show that we can use such partial models to compute minimal answers.  Firstly we reformulate Theorem 1.6 to give a characterisation of partial minimal answers in terms of cyclic strong covers.

\p

\noindent {\bf 3.3  Theorem.}    A set of predicates $\{A_i \vert i \le r\} $ is contained in a minimal answer iff for each $i\le r$ we may find a cyclic strong cover $\Cal C_i$ of $\{\lnot A_i\} \cup \{A_j \vert j\le r, j\ne i\}$ such that 
$T\models \bigvee_{i\le r} A_i \vee \bigvee \bigcap_{i\le r} \Cal C_i^+$.

\p

\noindent {\bf Proof}  ($\to$).    By Theorem 1.6, let $\{M_i \vert i\le r\}$ be a set of perfect models of $T$ such that for each $i\le r$,  $M_i\cap \{A_j \vert j \le r\}=\{A_i\}$, and $T\models \bigvee_{i\le r} A_i \vee \bigvee \bigcap_{i\le r} (\Cal L-M_i)$.  If $\Cal C_i = \overline{M_i} \cup (\Cal L - M_i)$, then it is easy to see that $\Cal C_i$ satisfies the required conditions.

\noindent $(\leftarrow$).    By Theorem 3.2, we may, for each $i$, pick a perfect  model $M_i$ of $T$ such that $M_i\not\models \bigvee \Cal C_i$.   It is then easy to check that the models $M_i$ satisfy the conditions of Theorem 1.6.   \bb

\p

We will now  use the above result to construct minimal answers.   At each stage in our construction we will generate a sequence of predicates $(A_1, A_2, \dots, A_r)$ which is witnessed (as representing a partial minimal answer) by a corresponding sequence of cyclic strong covers $(\Cal C_1, \Cal C_2, \dots, \Cal C_r)$ satisfying the conditions of Theorem 3.3.   

Given such a partial answer (and witnessing  cyclic strong covers) $((A_i, \Cal C_i) \vert i \le r))$, we then seek to extend this to a sequence of the form $((A_i,\Cal D_i) \vert  i \le r+1)$ satisfying the conditions of Theorem 3.3, and for which $\Cal D_i\supseteq \Cal C_i$ for each $i\le r$.      This extension will take place via two distinct phases.

Firstly we pick $A_{r+1}$ such that   $\{\lnot A_{r+1}\}\cup \{A_i \vert i\le r\}$ has a cyclic strong cover $\Cal F_{r+1}$ and for each $i\le r$,  $\{A_{r+1}\}\cup \Cal C_i$ has a cyclic strong cover $\Cal F_i$.   Note that $((A_i, \Cal F_i) \vert i \le r)$ continues to satisfy the conditions of Theorem 3.3 (since each $\Cal F_i\supseteq \Cal C_i$), but that  $((A_i, \Cal F_i) \vert i \le r+1)$ may not, since there is no guarantee that $T\models \bigvee_{i\le r+1} A_i  \vee \bigvee \bigcap_{i\le r+1} \Cal F_{i+1}^+$.   

The second phase thus attempts to find (for each $i\le r+1$) a cyclic strong cover $\Cal D_i$ of  $\Cal F_i$  such that $T\models \bigvee_{i\le r+1} A_i  \vee \bigvee \bigcap_{i\le r+1} \Cal D_i^+$, and it is the existence of the sets $\{\Cal D_i \vert i \le r+1\}$ which {\it verifies} the choice of $A_{r+1}$ and $\{\Cal F_i \vert i \le r+1\}$  made in the first phase.      If such verification is not possible, then we {\it truncate}  $((A_i, \Cal F_i) \vert i \le r+1)$ in order to seek alternative extensions of $((A_i, \Cal F_i) \vert i \le r)$.

The following definition captures these ideas.

\p

\noindent {\bf 3.4  Definitions.} 

\item{\bf (a)}     A {\it cyclic state} of length $r$ ($r\ge 1$) is a sequence $\Cal S = ((A_i, \Cal C_i) \vert i \le r)$ such that \newline  {\bf (i)}  each $A_i$ is a predicate, 
\newline 
{\bf (ii)}  each $\Cal C_i$
 is a cyclic strong cover of $\{\lnot A_i\}\cup \{A_j \vert j\le r, j\ne i\}$,   and
\newline {\bf (iii)}  for $1\le j<r$, $T\models \bigvee_{i\le j} A_i \vee \bigvee \bigcap_{i\le j} \Cal C_i^+$. 

\item {\bf (b)}    A cyclic state $\Cal S^* = ((A_i, \Cal D_i)\vert i \le r+k)$ ($k\ge 0)$ is an {\it extension} of $\Cal S = ((A_i, \Cal C_i)\vert i \le r)$ iff $\Cal C_i\subseteq \Cal D_i$ for each $i\le r$.

\item {\bf (c)}  A cyclic state  $\Cal S = ((A_i, \Cal C_i) \vert i \le r)$  is said to be:
\newline {\bf (i)}  {\it verified} iff $T\models \bigvee_{i\le r} A_i \vee \bigvee \bigcap_{i\le r} \Cal C_i^+$ (cf., Theorem 3.3), 
\newline {\bf (ii)}   {\it verifiable} iff $\Cal S$ has a verified extension,
\newline {\bf (iii)}  {\it total} iff each $\Cal C_i$ is total, and 
\newline {\bf (iv)}   {\it complete} iff $T\models \bigvee_{i\le r} A_i$.

\p

If $((A_i, \Cal C_i) \vert i \le r)$ is a cyclic state, note that $A_i\ne A_j$ for $i\ne j$ (since $\Cal C_i$ is consistent).   
Property (a)(iii) is a mere technical property  capturing our intention that a cyclic state will only be extended in length if it is verified.   This property also ensures that 
if $((A_i, \Cal C_i) \vert i \le r)$ is a cyclic state of length $r>1$, then truncation yields  a verified cyclic state $((A_i, \Cal C_i) \vert i \le r-1)$. 

Note the obvious fact that if, for each $i\le r$, $\Cal D_i$ is a cyclic strong cover of $\Cal C_i$, then $\Cal S^*=((A_i, \Cal D_i) \vert i \le r)$ is a cyclic state.  Moreover if $((A_i, \Cal C_i) \vert i \le r)$ is verified, then so is $\Cal S^*$.   The following proposition captures some of the basic properties of cyclic states.

\p

\noindent {\bf 3.5  Proposition.} 

\item {\bf (a)}    If $T\models \bigvee \{A_i \vert i \le r\}$, then $\{A_i \vert i \le r\}$  is a minimal answer iff there is a cyclic state of the form $((A_i, \Cal C_i)\vert i \le r)$.  

\item {\bf (b)}   If $T\models \bigvee \Cal A$, and for each $A\in \Cal A$,  $\Cal C_A$ is a cyclic strong cover of $\{\lnot A\}\cup (\Cal A-\{A\})$, 
then for any non-empty subset $\{A_1, A_2, \dots, A_r\}\subseteq \Cal A$, we have that $((A_i, \Cal C_{A_i}) \vert i \le r)$ is a verified cyclic state.    

\item {\bf (c)}  Every cyclic state of length 1 is verified.

\item {\bf (d)}  If $\Cal S = ((A_i, \Cal C_i) \vert i \le r)$  is verified, then $\Cal S$ has a complete total extension $((A_i, \Cal D_i) \vert i $ $\le r+k)$ such that $\{A_j \vert r<j\le r+k\}\subseteq \bigcap_{i\le r} \Cal C_i^+$.

\item {\bf (e)}   A set of predicates $\{A_i \vert i \le r\} $ is contained in a minimal answer iff there is a verified cyclic state of the form $((A_i, \Cal C_i)\vert i \le r)$.

\item {\bf (f)}  $\Cal S$ is verifiable iff $\Cal S$  has a verified extension of the same length iff $\Cal S$ has a complete extension iff $\Cal S$ has a total complete extension.

\p

\noindent {\bf Proof (a)}   This follows immediately from Theorem 3.3.

\noindent {\bf (b)}   This is trivial from the fact that if $j\le r$, then 
$\Cal A \subseteq \{A_i \vert i \le j\} \cup \bigcap_{i\le j} \Cal C_{A_i}^+$.

\noindent {\bf (c)}   A cyclic state of length 1 contains a single pair of the form $(A, \Cal C)$, where $\Cal C$ is a cyclic strong cover and $A\in \Cal C^-$, whence the result follows immediately from  Theorem 2.4(c).

\noindent {\bf (d)}    Let $\Cal A$ be a minimal answer with $\Cal A\subseteq \{A_i\vert i \le r\}\cup \bigcap_{i\le r} \Cal C_i^+$.   For each $i\le r$, $\Cal C_i$ is a cyclic strong cover of $\{\lnot A_i\}\cup (\Cal A - \{A_i\})$, whence $A_i\in \Cal A$.  For $i\le r$,  let $\Cal D_i$ be a total cyclic strong cover of $\Cal C_i$.

If we write $\Cal A = \{A_i \vert i \le r+k\}$, and for each $i>r$, let $\Cal D_i$ be a total cyclic strong cover of $\{\lnot A_i\}\cup (\Cal A - \{A_i\})$, then $((A_i, \Cal D_i) \vert i \le r+k)$ is the required cyclic state.

\noindent {\bf (e)}   If $\{A_i \vert i \le r+k\}$ is a minimal answer, then for each $i\le r+k$ we may find a total cyclic strong cover $\Cal C_i$ of $\{\lnot A_i\}\cup \{A_j \vert j \le r+k, j\ne i\}$.  By part (b),  $((A_i, \Cal C_i) \vert i \le r)$ is verified.   

 If $((A_i, \Cal C_i) \vert i \le r)$ is a verified cyclic state, then $\{A_i\vert i \le r\}$ is contained in a minimal answer by part (d).

\noindent {\bf (f)}  If         $\Cal S=((A_i, \Cal C_i) \vert i \le r)$  is verifiable, then we may find a verified extension $\Cal S'=((A_i,\Cal C_i')\vert i \le r+k)$.   If $k>0$, then by property 3.4(a)(iii), $S^*=((A_i,\Cal C_i')\vert i \le r)$ is the required verified extension.    By part (d),  $\Cal S'$ has a total complete extension.   

The converse follows trivially from part (b).   \bb

\p

We now present the notion  an  {\it immediate extension}, this being our mechanism  that will be used to extend cyclic states.     We require that our extension mechanism is correct, in the sense that repeated application  should generate only partial answers, and this will be guaranteed  by Theorem 3.3 and the use of verification.

We also require  completeness, meaning  that every minimal answer can be generated via repeated application of our extension mechanism.   Now every minimal answer is witnessed by a total complete cyclic state,  thus in order to achieve completeness we will insist that 
whenever   $\Cal S^*$ is a total complete extension of $\Cal S$, there is a (proper)  immediate extension $\Cal S'$ of $\Cal S$ such that $\Cal S^*$ is an extension of $ \Cal S'$.  The minimal answer represented by $\Cal S^*$ can then be constructed from $\Cal S$ via a sequence of immediate extensions.

Of course initially we start our computation of minimal answers with the empty sequence, whose immediate extensions we take to be of the form $((A, \Cal C))$, where $\Cal C$ is a constructible extension of $\{\lnot A\}$ (Definition 2.11).

\p

\noindent {\bf 3.6 Extending unverified cyclic states.}

Suppose now that $\Cal S = ((A_i, \Cal C_i)\vert i \le r)$ is an unverified cyclic state.   Our aim is to find cyclic strong covers $\Cal D_i$ of $\Cal C_i$ such that $T\models \bigvee_{i\le r} A_i \vee \bigvee \bigcap_{i\le r} \Cal D_i^+$.   There are broadly two ways in which we can extend the sets $\Cal C_i$:  either by adding some negative literal to some $\Cal C_{i_0}$ (which in turn will probably require the    addition of further literals (both positive and negative) in order to re-form a cyclic strong cover); or the addition of some positive literal (i.e., predicate) to some $\Cal C_{i_0}$ (which again would probably require the    addition of further literals).    However, in order to achieve our desired goal, we specifically  need to extend $\bigcap_{i\le r} \Cal C_i^+$, and this suggests that searching for positive literals which can be used to simultaneously extend all $\Cal C_i$ would be the more fruitful option.   This option is also  more appealing since it is the more constrained (i.e., has a narrower search space), and also because it allows within it an integral  test to check whether or not the current cyclic state is indeed unverified.   The following result encapsulates these ideas, and also shows that our approach to extending unverified cyclic states satisfies the completeness criteria above.

\p

\noindent {\bf 3.6.1  Lemma.}  Suppose that $\Cal S = ((A_i, \Cal C_i)\vert i \le r)$ is unverified, and $\Cal C$ is a cyclic strong cover of $\{A_i \vert i \le r\} \cup \bigcap_{i\le r}  \Cal C_i^+$.    Suppose also that $\Cal S$ is verifiable, and that $((A_i, \Cal D_i) \vert i \le r$) is a verified cyclic state such that for each $i\le r$, $\Cal D_i\supseteq \Cal C_i$.

Then $\Cal F = \bigcap_{i\le r} \Cal D_i^+ - ( \Cal C^+\cup \bigcup_{i\le r}  \Cal C_i^-) \ne \emptyset$, and for each $A\in \Cal F$ and each $i \le r$ we may find a constructible extension $\Cal C_i^*$ of $\{A\}\cup \Cal C_i$ such that 
$\Cal C_i^*\subseteq \Cal D_i$.

\p

\noindent {\bf Proof.}  First note that $\Cal C^+ \not\supseteq \bigcap_{i\le r} \Cal D_i^+ $, for otherwise $\Cal C$ would be a cyclic strong cover of $\{A_i \vert i \le r\}\cup \bigcap_{i\le r} \Cal D_i^+$, thus contradicting the fact that $T\models \bigvee_{i\le r} A_i \vee \bigvee \bigcap_{i\le r} \Cal D_i^+$.  
If $A\in \bigcap_{i\le r} \Cal D_i^+$, then  $A\not\in \bigcup_{i\le r} \Cal D_i^- \supseteq \bigcup_{i\le r}\Cal C_i^-$, and hence $\Cal F \ne \emptyset$.  

Finally, if $i \le r$ then   $\{A\}\cup \Cal C_i\subseteq \Cal D_i$, and the existence of $\Cal C_i^*$ is then given by Corollary 2.12.    \bb 

\p

This then allows us to define immediate extensions of unverified cyclic states.

\p

\noindent {\bf 3.6.2  Definition.}

Suppose that we are given a cyclic state $\Cal S = ((A_i, \Cal C_i) \vert i \le r)$, then an {\it immediate  extension} $\Cal S^*$ of $\Cal S$ is formed as follows.                                             
Pick a constructible extension $\Cal C$ of 
$\{A_i \vert i \le r\}\cup \bigcap_{i\le r}  \Cal C_i^+$.    (If no such $\Cal C$ exists, i.e., $\Cal  S$ is verified, then  immediate extensions of $\Cal S$  are as given in  Definition  3.7.3 below.)  

Pick a predicate $A\in \Cal L  - (\Cal C^+\cup \bigcup_{i\le r}  \Cal C_i^-)$ such that for each $i \le r$, $\{A\}\cup \Cal C_i$ has a constructible extension $\Cal C_i^*$, and let $\Cal S^* = ((A_i, \Cal C_i^*) \vert i \le r)$.    

If no such predicate $A$ exists, i.e., $\Cal S$ is not verifiable, then $\Cal S$ has no immediate extension and  the {\it truncation} of $\Cal S$ is given by $\Cal S^*=((A_i, \Cal C_i) \vert i \le r-1$).

\p

Note that if the predicate $A$ exists, then the extension formed is a proper extension, since  $A\in \Cal L - \Cal C^+\subseteq \Cal L - \bigcap_{i\le r} \Cal C_i^+$.     

The constraint that $A\in \Cal L  - (\Cal C^+\cup \bigcup_{i\le r}  \Cal C_i^-)$ clearly allows us to limit the search space, but nevertheless, it is still the case that we are {\it blindly} picking an element of $\Cal L  - (\Cal C^+\cup \bigcup_{i\le r}  \Cal C_i^-)$, and then determining whether the sets $\Cal C_i^*$ exist.   Note however that if we make an incorrect choice for $A_{r+1}$ then the cost of doing so is the cost of computing some of the sets $\Cal C_i^*$.     In fact this computation is not totally wasted, and we return to this point in the notes following Definition 3.7.3.    In Section 5 we will see that compilation allows us to partially overcome this need to make a blind choice.

Constructible extensions $\Cal C$ of 
$\{A_i \vert i \le r\}\cup \bigcap_{i\le r}  \Cal C_i^+$ can be computed with no additional effort by computing constructible extensions $\Cal C'$ of $\{A_i \vert i \le r\}$, and then computing constructible extensions $\Cal C$ of $\Cal C'\cup \bigcap_{i\le r}  \Cal C_i^+$.    This gives us a free useful test, since (by Theorem 3.2) if no such $\Cal C'$ exists, then $T\models \bigvee_{i\le r} A_i$, whence  $\{A_i\vert i \le r\}$ is a minimal answer and no further extension steps are required.   

This partitioning of the computation of constructible extensions of $\{A_i \vert i \le r\}\cup \bigcap_{i\le r}  \Cal C_i^+$ is also beneficial due to the fact that if we are successful in extending $((A_i, \Cal C_i) \vert$ $ i \le r)$ to a verified cyclic state $\Cal S' = ((A_i, \Cal D_i) \vert i \le r)$, then the subsequent extension  of $\Cal S'$ will again require the use (or computation) of the  cyclic strong covers of $\{A_i\vert i \le r\}$.   This is illustrated in the following section.

\p

\noindent {\bf 3.7 Extending verified cyclic states.}

 Suppose that $\Cal S = ((A_i, \Cal C_i) \vert i \le r)$ is verified (but $T\not\models \bigvee_{i\le r} A_i$, i.e., $\{A_i \vert i \le r\}$ has a cyclic strong cover).   An extension of $\Cal S$ will have the form $\Cal S^* = ((A_i, \Cal D_i) \vert i \le r+1)$, where (i) for $i\le r$,  $\Cal D_i$ is a cyclic strong cover of $\Cal C_i\cup \{A_{r+1}\}$, and (ii)  $\Cal D_{r+1}$ is a cyclic strong cover of $\{\lnot A_{r+1}\}\cup \{A_i\vert i\le r \}$.   

Cyclic strong covers of $\{\lnot A_{r+1}\}\cup \{A_i\vert i\le r \}$ can be computed by first computing a constructible extension  $\Cal C$ of $\{A_i\vert i\le r \}$, and then extending to a constructible extension  of $\{\lnot A_{r+1}\}\cup \Cal C$.    We thus have the choice of either guessing $A_{r+1}$ first, and then attempting to find the sets $\Cal D_i$ (if such exist), or choosing $\Cal C$ first, and then attempting to find $A_{r+1}$.  The latter option is the more fruitful for three  reasons:    Firstly, we have already computed the constructible extensions of $\{A_i \vert i \le r\}$ above.  Secondly,   given $\Cal C$ we can (by Lemma 3.7.1 below) prune the search space by insisting that 
$A_{r+1}\in \Cal L - (\Cal C^+\cup \bigcup_{i\le r} \Cal C_i^-)$, and finally given $\Cal C$, a predicate $A_{r+1}$ together with sets $\Cal D_i$ ($i\le r+1$) are guaranteed to exist (Lemma 3.7.2 below).   Lemma 3.7.1 shows that this strategy for extending $\Cal S$ satisfies our completeness condition identified in the remarks following Proposition 3.5.

\p

\noindent {\bf 3.7.1  Lemma.} Suppose that  $\Cal S = ((A_i, \Cal C_i) \vert i \le r)$ is a verified cyclic state, and that $((A_i, \Cal F_i) \vert i \le r+k)$ is a complete extension of $\Cal S$ with $k>0$.

Then we may find  a constructible extension $\Cal C$ 
of $\{A_i \vert i\le r\}$ such that 
\item{\bf (i)}  $A_{r+1}\not \in \Cal C^+\cup \bigcup_{i\le r} \Cal C_i^-$, 
\item{\bf (ii)}   $\{\lnot A_{r+1}\}\cup \Cal C$ has a constructible extension $\Cal D_{r+1}$ such that $\Cal D_{r+1}\subseteq \Cal F_{r+1}$, and 
\item{\bf (iii)}  for each $i \le r$,  $\{A_{r+1}\}\cup \Cal C_i$ has a  constructible extension  $\Cal D_i$  such that $\Cal D_i\subseteq \Cal F_{i}$.

\p

\noindent {\bf Proof.}    Trivially $A_{r+1}\in \bigcap_{i\le r}  \Cal F_{i}^+  \subseteq  \Cal L - \bigcup_{i\le r} \Cal F_{i}^-  \subseteq\Cal L - \bigcup_{i\le r} \Cal C_{i}^-   $.  Now $\Cal F_{{r+1}}$ is a  cyclic strong cover of $\{\lnot A_{r+1}\}\cup \{A_i \vert i \le r\}$, whence by Corollary 2.12 we may find a  constructible extension $\Cal C$ of $\{A_i \vert i \le r\} $ such that $\Cal C \subseteq \Cal F_{{r+1}}$.  Since $\lnot A_{r+1}\in \Cal F_{{r+1}}$, we must have that $A_{r+1}\not \in \Cal C^+$.

Again using Corollary 2.12, parts (ii) and (iii) follow trivially from the  facts that $\Cal F_{{r+1}}$ is a cyclic strong cover of $\{\lnot A_{r+1}\}\cup \Cal C$, and (for each $i\le r$)  $\Cal F_{i}$ is a cyclic strong cover of $\{A_{r+1}\}\cup \Cal C_i$.  \bb

\p

\noindent {\bf 3.7.2  Lemma.}      Suppose that  $\Cal S = ((A_i, \Cal C_i) \vert i \le r)$ is a verified cyclic state, and that  $\Cal C$ is a  cyclic strong cover 
of $\{A_i \vert i\le r\}$.  Then we may find  a predicate $A_{r+1}\in \Cal L - (\Cal C^+\cup \bigcup_{i\le r} \Cal C_i^-)$ such that 
\item{\bf (i)}   $\{\lnot A_{r+1}\}\cup \Cal C$ has a constructible extension $\Cal D_{r+1}$, and 
\item{\bf (ii)}  for each $i \le r$,  $\{A_{r+1}\}\cup \Cal C_i$ has a  constructible extension  $\Cal D_i$.

\p

\noindent {\bf Proof.}   For each $i\le r$, let $\Cal F_i$ be a total cyclic strong cover of $\Cal C_i$, and 
let $\Cal F$ be a total cyclic strong cover of  $ \Cal C$.   

Since $T\models \bigvee\{A_i\vert i \le r\}\vee \bigvee\bigcap_{i\le r} \Cal F_i^+$ we cannot have that $\Cal F\supseteq \{A_i\vert i \le r\}\cup \bigcap_{i\le r} \Cal F_i^+$, whence pick $A_{r+1}\in \{A_i\vert i \le r\}\cup \bigcap_{i\le r} \Cal F_i^+ - \Cal F$.    But then $A_{r+1}\not\in \Cal C^+\supseteq \{A_i\vert i \le r\}$, whence for each $i\le r$, $A_{r+1}\in \Cal F_i^+\subseteq \Cal L - \Cal F_i^-\subseteq \Cal L - \Cal C_i^-$.  In addition, since $\Cal F$ is total, $\lnot A_{r+1}\in \Cal F$.  

Parts (i) and (ii) then follow from Corollary 2.12 and the facts that $\{\lnot A_{r+1}\}\cup \Cal C\subseteq \Cal F$ and $\{A_{r+1}\}\cup \Cal C_i\subseteq \Cal F_i$.   \bb

\p

This then gives us our method of extending $((A_i, \Cal C_i) \vert i \le r)$. 

\p

\noindent {\bf 3.7.3  Definition.}   

Let $\Cal S = ((A_i, \Cal C_i) \vert i \le r)$ be a verified cyclic state,  then an {\it immediate  extension} $\Cal S^*$ of $\Cal S$ is formed as follows.  
Let $\Cal C$ be a constructible extension of $\{A_i \vert i \le r\}$.  (If no such $\Cal C$  exists then $T\models \bigvee_{i\le r} A_i $ and $\bigvee_{i\le r} A_i $ is a minimal answer.)

Pick $A_{r+1}\in \Cal L - (\Cal C^+\cup \bigcup_{i\le r} \Cal C_i^-$) such that \item{\bf (i)}   $\{\lnot A_{r+1}\}\cup \Cal C$ has a constructible extension $\Cal D_{r+1}$, and 
\item{\bf (ii)}  for each $i \le r$,  $\{A_{r+1}\}\cup \Cal C_i$ has a  constructible extension  $\Cal D_i$.

Let $\Cal S^* = ((A_i, \Cal D_i) \vert i \le r+1)$.

\p

\noindent {\bf Notes.}

\noindent {\bf 1.}    Note in both Definitions 3.6.2 and 3.7.3, that cyclic strong covers of $\{K\}\cup \Cal D$ could  be computed from constructible extensions of $\{K\}$ in the reduced database $T_{\Cal D}$ (by Theorem 2.9).  In the case when $K=\lnot A_{r+1}$ (Definition 3.7.3(i)) it may be preferable to adopt a different strategy (see 3 below).

\noindent {\bf 2.}     As mentioned in Section 3.6, cyclic strong covers may be computed and then found to be obsolete as far as  extending the current partial answer is concerned.    Note however 
that the computation of any cyclic strong cover $\Cal C$ is not wasted (provided $\Cal C^-\ne\emptyset$) since it can still  be employed in the derivation of other minimal answers:  for each $A\in \Cal C^-$, (($A, \Cal C))$ is a (verified) cyclic state of length 1.

\noindent {\bf 3.}  
Again we see that the constraint $A_{r+1}\in \Cal L - (\Cal C^+\cup \bigcup_{i\le r} \Cal C_i^-$) allows us to prune the search space, but that beyond this we are still making a blind choice of $A_{r+1}$, and then testing whether it satisfies conditions (i) and (ii) above.     This blind search is again not so detrimental, since it can provide useful information:   A constructible extension $\Cal D_{r+1}$ of   $\{\lnot A_{r+1}\}\cup \Cal C$ may be computed by first computing  a constructible extension $\Cal C'$ of $\{\lnot A_{r+1}\}$, and then computing $\Cal D_{r+1}$ as a constructible extension of $\Cal C'\cup \Cal C$.   In the case when no such $\Cal C'$ exists, we can immediately discount  $A_{r+1}$ (from belonging to any minimal answer).   In the case when $\Cal C'$ exists, but cannot be extended to $\Cal D_{r+1}$, then as above, for each $A\in \Cal C'^-$, the pair $(A, \Cal C')$ can still be employed in the derivation of other minimal answers, and the computation of $\Cal C'$ is not wasted.

\noindent {\bf 4.}    If $\Cal S = ((A_i, \Cal C_i) \vert i \le r)$   is verified, then by Proposition 3.5(d)  we may find a minimal answer $\Cal A$ such that $\{A_i \vert i \le r\}\subseteq \Cal A \subseteq \{A_i \vert i \le r\}\cup \bigcap_{i\le r} \Cal C_i^+$.   Moreover, it is easy to amend the proof of Lemma 3.7.2 to show that given a constructible extension $\Cal C$ of $\{A_i\vert i \le r\}$ we can find some $A_{r+1}$ in $\Cal A - \{A_i \vert i\le r\}$ (and hence in $\bigcap_{i\le r} \Cal C_i^+$) satisfying conditions (i) and (ii) of Definition 3.7.3.   
\newline \indent Note however that in Definition 3.7.3 we are {\it not} able to insist that $A_{r+1}$ is chosen from $\bigcap_{i\le r} \Cal C_i^+$, since this would compromise completeness.    For example if $T=\{A\vee B, A\to C, B\to D\}$, then $C\vee D$ is  a minimal answer, but the only constructible extension of $\lnot C$ (resp. $\lnot D)$ is $\{\lnot C, \lnot A, B\}$ (resp. $\{\lnot D, \lnot B, A\}$), whence the only cyclic states of length 1 (generated using constructible extensions) representing a sub-answer of $C\vee D$ are $((C, \{\lnot C, \lnot A, B\}))$ and $((D, \{\lnot D, \lnot B, A\}))$.      \indent 
This unfortunate inability to further limit the search space could be overcome by allowing verified states to be extended in the manner similar to that suggested  in  Section 3.6.2, but then such a requirement effectively insists upon the computation of perfect models in their entirety.  We will see in Section 5 that compilation goes quite some way to overcoming this inability.

\noindent {\bf 5.}   We have already mentioned that our method is both complete and correct.  Notice that a truncation step is in effect an undo operation, and clearly we wish to prevent circularity by insisting that extension steps following truncation do not redo what has previously been undone.   With this proviso, it is then clear that any sequence of cyclic states generated via  immediate extension and truncation will eventually generate a complete cyclic state (i.e., a minimal answer).   

\p

\p

\noindent {\bf \S 4. \,\,  Unstratified databases}

\p

For unstratified databases, cyclic strong covers are not necessarily extendible to a disjunctive stable model, and we are thus (apparently) unable to characterise partial minimal answers without computing {\it total} cyclic strong covers (i.e., disjunctive stable models).  In addition, testing verification, i.e., whether $T\models \bigvee_{i\le r} A_i \vee \bigvee \bigcap_{i\le r} \Cal C_i^+$ (cf., Theorem 3.3) requires the computation of total cyclic strong covers  (Theorem 2.6(b)), at least in the case when the cyclic state in question is not verified.        

As discussed in [Jo99a] (and mentioned in Section 2 above), disjunctive stable models can be generated  by (implicitly) adding to our database, for each predicate $P\in \Cal L$, the denial rule $P\wedge \lnot P \to {\tt FALSE}$:     The disjunctive stable models of the database are unaffected, and every strong cover of $\{{\tt FALSE}\}$ is then total.      Cyclic states then encode sequences of disjunctive stable models satisfying the conditions of Theorem 1.6, and a (top-down) construction of cyclic strong covers [Jo98, Jo99, Jo99a] provides us with a means of testing verification.   The application of these denial rules within such  a top-down construction  amounts to the application of an unrestricted splitting rule (e.g., [Ya96]).    

As mentioned earlier, the computation of models in their entirety is undesirable, and in the following section we show that a partitioning of the database can be used to (partially) alleviate this need.

\p

\p

\noindent {\bf \S 5. \,\,   Compilation}

\p

It is  natural to ask  how our method compares in terms of computational efficiency with  methods that also generate non-minimal answers.   Is the checking required at each stage cost effective in relation to the saving (i.e., of not generating non-minimal answers)?   

In this section we show that compiling (pre-processing) the computation of cyclic strong covers can be employed to greatly  simplify and reduce the cost of the run-time computation.   In addition, as a by-product, compilation is shown to resolve the problems raised in Section 4 above.

Throughout this section we assume that $\Cal L$ is the disjoint union of EXT($\Cal L$) and INT($\Cal L)$. If $\Cal Q$ is a set of literals, then let $\Cal Q_{ext} = \{K\in \Cal Q \vert K\in \text{EXT}(\Cal L) \text{ or } \lnot K \in  \text{EXT}(\Cal L)\}$, and $\Cal Q_{int} = \{K\in \Cal Q \vert K\in \text{INT}(\Cal L) \text{ or } \lnot K\in \text{INT}(\Cal L)\} =  \Cal Q - \Cal Q_{ext}$.  Notice that $(\Cal Q^-)_{ext} = (\Cal Q_{ext})^-$, etc.   We  also make the (usual) assumption  that for each rule $C$, either 
\item{(i)}   conseq($C)\subseteq  \text{EXT}(\Cal L)$ and antec($C)\cup \Cal N(C) = \emptyset$, or
\item{(ii)}  conseq($C)\subseteq  \text{INT}(\Cal L)$ and antec($C)\cup \Cal N(C) \ne \emptyset$ (whence antec($C)\ne\emptyset$, cf., Section 1.1).

In case (ii) the assumption that the body of $C$ is non-empty if of course a technical requirement that can be achieved artificially without loss of generality.

We let   EXT($T) = \{C \in T \vert \,\,$conseq($C)\subseteq \text{EXT}(\Cal L)\}$, and INT($T)=T-\text{EXT}(T) = \{C\in T \vert \,\,$conseq($C)\subseteq \text{INT}(\Cal L)\}$.

Note that a rule in EXT($T$) has the form $\bigvee \Cal E$, where $\Cal E\subseteq \text{EXT}(\Cal L)$.   In particular,  minimal and disjunctive stable models of EXT($T$) coincide, thus if $\Phi$ is a formula in EXT($\Cal L$), then  EXT($T)\models \Phi$ iff $\Phi$ is true in every minimal model of $\Phi$.   In particular if $\Cal F\subseteq \text{EXT}(\Cal L)$, then EXT($T)\models \bigvee \Cal F$ iff there is a rule $\bigvee \Cal E\in \text{EXT}(T)$ such that $\Cal E \subseteq \Cal F$.
 
This partitioning of $\Cal L$ can be viewed as a very weak form of stratification, and indeed yields a weakened form of Theorem 3.2 as follows:   Let us say that a consistent set of literals $\Cal Q$ is {\it int-total} iff $\Cal Q^-\cup \Cal Q^+\supseteq \text{INT}(\Cal L)$.    If $\Cal C$ is an int-total  cyclic strong cover, then 
EXT($T)\not\models \bigvee \Cal C^+_{ext}$, whence we may find a minimal model $M_0\subseteq \text{EXT}(\Cal L)$ of EXT($T$) such that $M_0\cap \Cal C^+_{ext}=\emptyset$.  But then it is easy to show that $\Cal C_{ext}^-\subseteq M_0$, and hence that  $\Cal C\cup \overline{M_0}\cup (\text{EXT}(\Cal L)-M_0)$ is a total cyclic strong cover extending $\Cal C$.  This then yields the following result.

\p

\noindent {\bf 5.1 Theorem}  [Jo99a]. 

\item{\bf (a)}   Every int-total cyclic strong cover can be extended to a total cyclic strong cover.

\item{\bf (b)}   If  $\Cal Q$ is a  set of literals, 
then $T\models  \bigvee \Cal Q$ iff $\Cal Q$ has no int-total cyclic strong cover.

\p

Compilation is based upon the assumption that EXT($T$) is relatively transient, in contrast to INT($T$) which is assumed to be relatively static.   Compilation is then the pre-processing of INT($T$) so that subsequent run-time query processing requires a manipulation of EXT($T$) only.    Whilst (the less frequent) modifications to INT($T$) necessitate recompilation, the more frequent modifications to EXT($T$)  do not.    

Since query processing requires the construction of int-total cyclic strong covers, we need to partition this construction into a computation step against INT($T$), and then a further step against EXT($T$).   

For strong covers this partitioning is trivial, since a strong cover in $T$ is simply a set that is both a strong cover in INT($T$) and a strong cover in EXT($T$).   Moreover note that a consistent set of literals $\Cal C$ is a strong cover in EXT($T$) iff EXT($T)\not\models \bigvee \Cal C_{ext}^+$.

For cyclic trees, every leaf node is a rule node.  Moreover by assumptions (i) and (ii) above, a rule node $rn_C$ is a  leaf node iff $C\in \text{EXT}(T)$.   If $rn_C$ is such a leaf node, with parent $n$, then conseq($C)\cap \text{CYC}(n) \ne \emptyset$, whence the branch to $n$ must contain a predicate node labelled with an extensional predicate.  If $m$ is the top-most such predicate node, then CYC($m)=\{lab(m)\}$, whence the child of $m$ must be a rule node of the form $rn_D$, where conseq($D)\cap \text{CYC}(m)\ne\emptyset$, whence $D\in \text{EXT}(T)$, i.e., $m=n$.  Thus predicate nodes are labelled with extensional predicates iff their child is a leaf node.

Let us therefore say that a {\it partial cyclic tree} satisfies the conditions of Definition  2.3, with the exception that every leaf node is a predicate node labelled with an extensional  predicate.      Such trees are defined (constructed) entirely within INT($T$).    To {\it complete}  a partial cyclic tree in order to form a  cyclic tree, we need to extend each such leaf node  with a rule node $rn_C$, where $C\in \text{EXT}(T$).   In order to ensure that the extended tree continues to satisfy the conditions of Definition 2.3, we must have that $lab(n) \in \text{conseq}(C)$ (since CYC($n)=\{lab(n)\})$ and $\Cal O(rn_C) =\text{conseq}(C)-\{lab(n)\}$ is disjoint from $Pred(\Cal T)$.     The extended tree $\Cal T'$ then has the properties that $Pred(\Cal T') = Pred(\Cal T)$,  $\Cal N(\Cal T') = \Cal N(\Cal T)$ and $\Cal O(\Cal T') = \Cal O(\Cal T) \cup \bigcup \{\Cal O(rn_C) \vert rn_C \text{ is a leaf in } \Cal T'\}$.

\p

\noindent {\bf 5.2 Definition}  [Jo99, Jo99a].  A consistent set of literals $\Cal C$ is a {\it weakly cyclic cover} (in INT($T$)) iff
\item{\bf  (i)}  $\Cal C$ is a strong cover in INT($T$), and 
\item{\bf (ii)}  there is  a set $\{\Cal T_i \mid
i\le m\}$ of 
partial cyclic trees  such that 
$\Cal C^- = \bigcup_{i\le m} Pred(\Cal T_i)$, and 
$\Cal C^+ \supseteq \bigcup_{i\le m} (\Cal O(\Cal T_i) \cup \Cal
N(\Cal T_i))$.

Let $\Cal C$ be a weakly cyclic cover in INT($T$) and  $f$ be a function 
$f : \Cal C^-_{ext} \to 
\text{EXT}(T)$ such that for each $P\in \Cal C^-_{ext}$,  $f(P)=\bigvee \Cal E_P$, where  $P\in \Cal E_P$ and $(\Cal E_P-\{P\}) \cap \Cal C_{ext}^- = \emptyset$.    The set $\Cal C \cup \bigcup \{\Cal E_P - \{P\} \mid P \in 
\Cal C^-_{ext}\}$ is said
to be a {\it completion} of $\Cal C$ in $T$.   

\p

Thus a completion of a weakly cyclic cover is formed by completing each of the cyclic trees which form the weakly cyclic cover.    Note  that 
if $\Cal D$ is a completion of $\Cal C$, then $\Cal D - \Cal C
\subseteq \text{EXT}(\Cal L)$.    Note also 
that the computation of weakly cyclic covers takes place entirely in INT($T$), and the  computation of completions takes place entirely in EXT($T$).      

In fact it is easy to observe that a partial tree $\Cal T$ can always be extended to a cyclic tree in INT($T)\cup \{E \vert E \in Pred(\Cal T)_{ext}\}$ by appending to each predicate leaf node $n$ the rule node labelled with the unit rule $lab(n)$.   We can thus easily show that  a consistent set of literals $\Cal C$ is a  weakly cyclic cover (in INT($T$))  iff $\Cal C$ is a cyclic strong cover in 
INT($T) \cup \{E \vert  E \in \Cal C^-_{ext}\}$.      This then allows us to apply the previous results concerning cyclic strong covers to weakly cyclic covers.

Proposition 5.3 summarises the properties of weakly cyclic covers presented in [Jo99, Jo99a].

\p

\noindent {\bf 5.3 Proposition} [Jo99, Jo99a].

\item{\bf (a)}  Let $\Cal D$ be a cyclic strong cover.  Then  $\Cal D$ is a weakly cyclic cover, and moreover if $\Cal C$ is a weakly cyclic cover with $\Cal C \subseteq \Cal D$, there is a completion $\Cal C'$ of $\Cal C$ such that $\Cal C'\subseteq \Cal D$.

\item{\bf (b)}    If $\Cal C$ is a weakly cyclic cover and $\Cal D$ is a completion of $\Cal C$, then $\Cal D$ is a cyclic strong cover iff  EXT($T)\not\models \bigvee \Cal D_{ext}^+$. 

\item{\bf (c)}   If   $\Cal Q$ is a set of literals, then $T\models  \bigvee \Cal Q$ iff whenever $\Cal C$ is an int-total weakly cyclic cover of $\Cal Q$ and $\Cal D$ is a completion of $\Cal C$, then EXT($T)\models \bigvee \Cal D^+_{ext}$,

\p

Since weakly cyclic covers are characterised by INT($T$), we can easily show that if $\Cal Q$ is a set of literals, then $\Cal C$ is a weakly cyclic cover of $\Cal Q$ iff $\Cal C$ has the form $\Cal C=\Cal C'\cup \Cal Q_{ext}$ where $\Cal C'$ is a weakly cyclic cover of $\Cal Q_{int}$ and $\Cal C'\cup \Cal Q_{ext}$ is consistent.  This then yields the following  corollary.

\p

\noindent {\bf 5.4 Corollary.}    

\item{\bf (a)}   $\Cal D$ is a cyclic strong cover of $\Cal Q$ iff there is a weakly cyclic cover $\Cal C'$ of $\Cal Q_{int}$ such that $\Cal C'\cup \Cal Q_{ext}$ is consistent, and $\Cal D$ is a completion of $\Cal C'\cup \Cal Q_{ext}$ with  EXT($T)\not\models \bigvee \Cal D_{ext}^+$. 

\item{\bf (b)}   If   $\Cal Q$ is a set of literals, then    $T\models  \bigvee \Cal Q$ iff whenever $\Cal C'$ is an int-total weakly cyclic cover of $\Cal Q_{int}$ such that $\Cal C'\cup \Cal Q_{ext}$ is consistent, and $\Cal D$ is a completion of $\Cal C'\cup \Cal Q_{ext}$, then EXT($T)\models \bigvee \Cal D^+_{ext}$.

\p

\noindent {\bf 5.5   The compilation process.}

Clearly the compilation process requires the computation of int-total weakly cyclic covers.   As suggested in Sections 2 and  4, forcing weakly cyclic covers to be int-total can be simply achieved by implicitly adding $\{P\wedge \lnot P \to {\tt FALSE} \vert P \in \text{INT}(\Cal L)\}$ to INT($T$).    Note that during compilation we can (if desired) apply subsumption to remove redundancy, since if $\Cal C$ and $\Cal D$ are  int-total  weakly cyclic covers with $\Cal C \subseteq \Cal D$, then any completion of $\Cal D$ contains a completion of $\Cal C$.   An appropriate set of int-total weakly cyclic covers can thus be generated using a ``constructible'' approach (cf., Definition 2.11 and Corollary 2.12), the details of which are discussed further in [Jo98a, Jo99, Jo99a].

We assume therefore that the compilation process generates a set $\comp$ of int-total weakly cyclic covers, such that every int-total weakly cyclic cover is a superset of some element of $\comp$.  

\p

We now turn our attention to the computation of minimal answers following compilation.   Firstly note that cyclic strong covers can be computed using $\comp$ as follows.   If $\Cal Q$ is a consistent set of literals, and $\Cal S$ is an int-total cyclic strong cover of $\Cal Q$, then we may find some $\Cal C\in \comp$ such that $\Cal C\supseteq \Cal Q_{int}$,  $\Cal C\cup \Cal Q_{ext}$ is consistent, and a completion $\Cal D$ of $\Cal C \cup \Cal Q_{ext}$ such that $\Cal D \subseteq \Cal S$ (whence      EXT($T)\not\models \bigvee \Cal D_{ext}^+$ and $\Cal D$ is a cyclic strong cover of $\Cal Q$).    Let us denote by $\comp(\Cal Q)$ the set of cyclic strong covers of $\Cal Q$ so obtained, i.e., 

$\comp(\Cal Q) = \{\Cal D\vert \,\exists \Cal C \in \comp, \Cal C\supseteq \Cal Q_{int}, \Cal C\cup \Cal Q_{ext} \text{ is consistent},  \Cal D \text{ is a completion of }$\newline \p\hfill $\Cal C\cup \Cal Q_{ext}, \text{ and } \text{EXT}(T)\not\models \bigvee \Cal D_{ext}^+\}.$

As in Section 3, our run-time computation of minimal answers is again based upon two distinct processes, depending on whether the cyclic state to be extended is verified or not.

\p

\noindent {\bf 5.6 Extending unverified cyclic states.}

As in Section 3, our aim is to extend an unverified cyclic state $\Cal S = ((A_i, \Cal C_i) \vert i \le r)$ to a verified cyclic state  $\Cal S^* = ((A_i, \Cal D_i) \vert i \le r)$, thus again we need to extend $\bigcap_{i\le r}\Cal C_i^+$ by the simultaneous addition of some predicate(s) to each $\Cal C_i^+$.    Since each $\Cal C_i$ will already be int-total, such predicates must be taken from EXT($\Cal L)$.   With this in mind, note that if $\Cal C$ is an int-total cyclic strong cover,  and $\Cal B\subseteq \text{EXT}(\Cal L)$, then $\Cal B \cup \Cal C$ is a cyclic strong cover iff $\Cal B\cap \Cal C^-_{ext}=\emptyset$ and EXT($T)\not\models \bigvee \Cal B \vee \bigvee \Cal C_{ext}^+$.    

 As in Section 3.6 we first perform a test to ensure that $\Cal S$ is not verified, and then use the results of this test to constrain the search space.

\p

\noindent {\bf 5.6.1 Theorem.}  Let $\Cal S = ((A_i, \Cal C_i) \vert i \le r)$ be a cyclic state, where each $\Cal C_i$ is 
int-total, and $\Cal V = \{A_i \vert i \le r\} \cup \bigcap_{i\le r} \Cal C_i^+$.  Suppose   that $\Cal S$ is not verified and that $\Cal D\in \comp(\Cal V)$.   Suppose also that $\Cal S$ is verifiable, and that  $((A_i, \Cal D_i) \vert i \le r)$ is a verified cyclic state, with each $\Cal D_i\supseteq \Cal C_i$.  

Then we may find a rule  $\bigvee \Cal E \in \text{EXT}(T)$ such that for each $i\le r$, $\Cal E - \Cal D \subseteq \Cal D_i^+ - \Cal C_i^-$,   $\Cal E\cap (\Cal D - \Cal C_i)\ne\emptyset$ and  EXT($T)\not\models 
\bigvee (\Cal E - \Cal D) \vee \bigvee (\Cal C_i^+)_{ext}$.  

\p

\noindent {\bf Proof.}     Let $\Cal W =  \{A_i \vert i \le r\} \cup \bigcap_{i\le r} \Cal D_i^+$, then $\Cal V \subseteq \Cal W$,  $\Cal W_{int}=\Cal V_{int}$ and $\Cal W- \Cal V\subseteq \text{EXT}(\Cal L)$.  

Now $\Cal D$ is a completion of some $\Cal C\cup \Cal V_{ext}$, where $\Cal C$ is a weakly cyclic cover of $\Cal V_{int}=\Cal W_{int}$ and $\Cal C\cup \Cal V_{ext}$ is consistent.  

If $\Cal C \cup \Cal W_{ext}$ is inconsistent, then there is some $P\in \Cal C^-$ such that $P\in \Cal W_{ext}-\Cal V_{ext}$, and (by the definition of a completion) there is some rule $\bigvee \Cal E\in \text{EXT}(T)$ such that $P\in \Cal E$ and $\Cal E-\{P\}\subseteq \Cal D$, whence $\Cal E\subseteq \Cal D \cup (\Cal W_{ext}-\Cal V_{ext})$.   If $\Cal C \cup \Cal W_{ext}$ is consistent, then  $\Cal D \cup (\Cal W_{ext} - \Cal V_{ext})$ is a completion of $\Cal C \cup \Cal V_{ext}\cup (\Cal W_{ext} - \Cal V_{ext})=\Cal C \cup \Cal W_{ext}$, whence by Corollary 5.4(b), there is some  $\bigvee \Cal E\in \text{EXT}(T)$ such that $\Cal E \subseteq \Cal D\cup (\Cal W_{ext} - \Cal V_{ext})$.

But then  $\Cal E - \Cal D \subseteq (\Cal W_{ext} - \Cal V_{ext})^+\subseteq \bigcap_{i\le r}  \Cal D_i^+\subseteq \Cal L - \bigcup_{i\le r}  \Cal D_i^-\subseteq \Cal L - \bigcup_{i\le r}  \Cal C_i^-$.    Suppose that $\Cal E \cap (\Cal D - \Cal C_i)=\emptyset$, then $\Cal E \cap \Cal D \subseteq \Cal C_i$ and therefore $\Cal E = (\Cal E -\Cal D)\cup (\Cal E\cap \Cal D) \subseteq \Cal D_i\cup \Cal C_i\subseteq \Cal D_i$,  thus contradicting the fact that $\Cal D_i$ is a  strong cover.   Similarly EXT($T)\not\models 
\bigvee (\Cal E - \Cal D) \vee \bigvee (\Cal C_i^+)_{ext}$, since $(\Cal E - \Cal D) \cup \Cal C_i^+\subseteq \Cal D_i$.           \bb

\p

\noindent {\bf 5.6.2  Definition.}

Suppose that $\Cal S = ((A_i, \Cal C_i) \vert i \le r)$ is a cyclic state, where each $\Cal C_i$ is int-total.  
 An {\it immediate extension} $\Cal S^* = ((A_i, \Cal D_i) \vert i \le r)$ of $\Cal S$ is computed as follows.       
  Pick $\Cal D\in \comp(\{A_i \vert i \le r\} \cup \bigcap_{i\le r} \Cal C_i^+)$.     By Corollary 5.4, if no such $\Cal D$ exists, then $\Cal S$ is verified, in which case immediate extensions of $\Cal S$ are defined 
in Definition 5.7.3 below.

Pick $\bigvee \Cal E \in \text{EXT}(T)$ such that  for each $i\le r$, $(\Cal E-\Cal D)\cap \Cal C_i^-=\emptyset$,  $\Cal E\cap (\Cal D - \Cal C_i)\ne\emptyset$ and   EXT($T)\not\models 
\bigvee (\Cal E - \Cal D) \vee \bigvee (\Cal C_i^+)_{ext}$.     We then set $\Cal D_i = (\Cal E- \Cal D)\cup \Cal C_i$. 

Notice that there must be some $i_0$ for which 
$\Cal C_{i_0}\subset \Cal D_{i_0}$, for otherwise  $\Cal E-\Cal D\subseteq \bigcap_{i\le r} \Cal C_{i}^+$, thus  $\Cal E\subseteq \Cal D \cup (\Cal E-\Cal D)\subseteq \Cal D\cup \bigcap_{i\le r} \Cal C^+_i \subseteq \Cal D$,  contradicting the fact that EXT($T)\not\models \bigvee \Cal D^+_{ext}$.  Thus  $\Cal S^*$ is a proper extension of $\Cal S$.

If no  such rule $\bigvee \Cal E$ exists in EXT($T$) then  $\Cal S$ is not verifiable (by Theorem 5.6.1), $\Cal S$ has no immediate extension, and the truncation of $\Cal S$ is given by $((A_i, \Cal C_i) \vert i \le r-1)$. 

\p

Theorem 5.6.1 demonstrates that our   extension step satisfies the required completeness property (given after Proposition 3.5). 

\p

\noindent {\bf 5.7 Extending verified cyclic states.}

We first present two results (whose proofs are virtually identical to those of Lemmas 3.7.1 and 3.7.2) detailing the extension mechanism for verified cyclic states and showing that it satisfies the required completeness property.

\p

\noindent {\bf 5.7.1  Lemma.} Suppose that  $\Cal S = ((A_i, \Cal C_i) \vert i \le r)$ is a verified cyclic state such that each $\Cal C_i$ is int-total, and let $((A_i, \Cal F_i) \vert i \le r+k)$  be a complete total extension of    $\Cal S$ with $k>0$.   Then we may find  a $\Cal C\in \comp(\{A_i \vert i\le r\})$ such that  
\item{\bf (i)}  $A_{r+1}\not \in \Cal C^+\cup \bigcup_{i\le r} \Cal C_i^-$,  
\item{\bf (ii)}   there is a $\Cal D_{r+1} \in \comp(\{\lnot A_{r+1}\}\cup \Cal C)$ such that $\Cal D_{r+1}\subseteq \Cal F_{{r+1}}$, and 
\item{\bf (iii)}  for each $i \le r$,  there is a $\Cal D_i \in \comp(\{A_{r+1}\}\cup \Cal C_i)$ such that $\Cal D_i\subseteq \Cal F_{i}$.

\p

\noindent {\bf 5.7.2  Lemma.}      Suppose that  $\Cal S = ((A_i, \Cal C_i) \vert i \le r)$ is a verified cyclic state where each $\Cal C_i$ is int-total, and that  $\Cal C\in \comp(\{A_i \vert i\le r\})$.  Then we may find  a predicate $A_{r+1}\in \Cal L - (\Cal C^+\cup \bigcup_{i\le r} \Cal C_i^-)$ such that 
\item{\bf (i)}   $\{\lnot A_{r+1}\}\cup \Cal C$ has an int-total cyclic strong cover $\Cal D_{r+1}$, and 
\item{\bf (ii)}  for each $i \le r$,  $\{A_{r+1}\}\cup \Cal C_i$ has an int-total  cyclic strong cover $\Cal D_i$.

\p

In Lemma 5.7.2, note that when $A_{r+1}\in \text{INT}(\Cal L)$, we must have that  $A_{r+1}\in \Cal C^-\cap \bigcap \Cal C_i^+$ (since $\Cal C$ and each $\Cal C_i$ is int-total), in which case $\Cal C$ itself is an int-total cyclic strong cover of  $\{\lnot A_{r+1}\}\cup \Cal C$, and for each $i\le r$, $\Cal C_i$ is itself an int-total cyclic strong cover of $\{A_{r+1}\}\cup \Cal C_i$.

In the case when $A_{r+1}\in \text{EXT}(\Cal L)$, condition (i) can be characterised by the existence of a rule $\bigvee \Cal E \in \text{EXT}(\Cal L)$ such that $A_{r+1}\in \Cal E$, $(\Cal E - \{A_{r+1}\})\cap \Cal C^-_{ext}=\emptyset$  and EXT($T)\not\models \bigvee (\Cal E - \{A_{r+1}\})\vee \bigvee \Cal C_{ext}^+$, whence we may take $D_{r+1} = \{\lnot A_{r+1}\}\cup (\Cal E - \{A_{r+1}\}) \cup \Cal C$ (i.e., $\Cal D_{r+1}$ is a completion of $\{\lnot A_{r+1}\}\cup  \Cal C$).   Condition (ii) can be characterised by the condition  EXT($T)\not\models A_{r+1}\vee \bigvee (\Cal C_i^+)_{ext}$, in which case  we may take $\Cal D_i = \{A_{r+1}\}\cup \Cal C_i$. As mentioned earlier, in both cases, the test against EXT($T$) is of course a simple subsumption check.

Note that in both cases, $A_{r+1}\not\in \{A_i \vert i \le r\}$ since $A_{r+1}\not \in \Cal C^+ \supseteq \{A_i \vert i \le r\}$.

\p

\noindent {\bf 5.7.3  Definition.}

Suppose that $\Cal S = ((A_i, \Cal C_i) \vert i \le r)$ is a 
verified cyclic state, where each $\Cal C_i$ is int-total.     An {\it immediate extension} $\Cal S^*$ of $\Cal S$ is computed as follows.    Let $\Cal C\in \comp(\{A_i \vert i \le r\})$.     (As in Section 3.7.3,  if no such $\Cal C$ exists, then $T\models \bigvee_{i\le r} A_i$ and  $\bigvee_{i\le r}A_i$ is a minimal answer.)    There are now two cases (at least one of which must apply), depending on whether we (try to) extend with a predicate  in INT($\Cal L$) or EXT($\Cal L$).

\item{\bf (a)}   Pick a predicate $A_{r+1}  \in \text{INT}(\Cal L) \cap \Cal C^- \cap \bigcap_{i\le r} \Cal C_i^+$, and let $\Cal S^*$ be formed by extending $\Cal S$ with the pair $(A_{r+1}, \Cal C)$.

\item {\bf (b)}  Pick a predicate $A_{r+1}\in  \text{EXT}(\Cal L) - (\Cal C^+\cup \bigcup_{i\le r}\Cal C_i^-)$ such that \newline {(i)} \phantom{i}for each $i \le r$,  EXT$(T)\not\models A_{r+1}\vee \bigvee (\Cal C_i^+)_{ext}$, and \newline 
{(ii)}   there is a rule $\bigvee \Cal E\in \text{EXT}(T)$ such  that $A_{r+1}\in \Cal E$,  $(\Cal E - \{A_{r+1}\})\cap \Cal C^-_{ext}=\emptyset$  and  EXT($T)\not\models \bigvee(\Cal E-\{A_{r+1}\})\vee\bigvee \Cal C_{ext}^+$.  \newline  $\Cal S^*$ is then formed from $\Cal S$ by replacing each $\Cal C_i$ by $\{A_{r+1}\}\cup \Cal C_i$, and then extending with the pair $(A_{r+1}, \{\lnot A_{r+1}\}\cup (\Cal E-\{A_{r+1}\})\cup \Cal C)$.

\p

\noindent {\bf 5.8 Compilation for query processing.}

Compilation has been previously studied for query processing under the minimal model [Ya02], perfect model [Jo99], disjunctive stable model [Jo99a], possible model [Jo02], and disjunctive well-founded [Jo03] semantics.     Each of these approaches compiles a specific query (as opposed to using compilation to generate all (minimal) answers).  For example in order to compile a query $\bigvee \Cal Q$ under the disjunctive stable model semantics, the approach presented in [Jo99a] computes int-total 
weakly cyclic covers $\Cal C$ of $\Cal Q$, and the run-time processing then attempts to find a completion $\Cal D$ of some such $\Cal C$ for which EXT($T)\not\models \bigvee \Cal D_{ext}^+$.   (In the terminology of the current section, $T\models \bigvee \Cal Q$ iff $\comp(\Cal Q)=\emptyset$.)

Compilation of the GCWA in first order non-recursive positive databases is discussed in [He88].   In such databases, evaluation under the GCWA reduces to testing minimal answer membership, whose compilation is achieved by a combination of resolution and subsumption.    

\p

\noindent {\bf 5.9 Other forms of pre-processing.}

Compilation as presented above is not the only form of pre-processing that can be applied.   For example, in our approach we have partitioned the database into two components, the first of which, EXT($T$), consists of simple disjunctions.  Our crucial result, Theorem 5.1, was dependent on the resulting properties of EXT($T$), specifically that testing whether EXT($T)\models \bigvee \Cal Q_{ext}$ can be achieved without recourse to cyclic strong covers that are total in EXT($\Cal L$).   But this property is (by Theorem 3.2) shared by all stratified databases, and we therefore note that the techniques of the current section can be extended to the case where the database is suitably partitioned into two components, one of which is stratified.    This approach was discussed briefly in the context of query processing in [Jo99a, Section 6].

An alternative, which is explored in [Jo99] for stratified databases, is to employ cyclic trees in order to transform a database into an equivalent  database which has specific properties.   For example we show in Appendix B that we can transform a database into a positive database which is equivalent under the disjunctive stable model semantics, and in which we can compute minimal answers of the original program.

As noted in the introduction,  a form of pre-processing referred to as {\it unfolding} has been studied in [Bra94, Bra95] in relation to a wide class of semantics.   This pre-processing transforms the database into a set of {\it conditional facts} (i.e., rules for which antec($C)=\emptyset$).   Such facts are of course closely related to answers, and in this respect one might argue that unfolding is closer to full scale answer generation than it is to pre-processing.      More importantly (from the database perspective), such a transformation could not be immune to changes in EXT($T$).   In [Jo99, Section 4],  cyclic trees are employed to transform a stratified database into an equivalent database  whose rules have the property that antec($C)\subseteq \text{EXT}(\Cal L)$.  Such a transformation is immune to changes in EXT($T$), although whether this result can be extended to the disjunctive stable model semantics is unclear.

\p

\p

\noindent {\bf \S 6. \,\,  Query answers}

\p

Let us say that a {\it query} to a database is an expression of the form ?$\bigvee \Cal H$, where $\Cal H \subseteq \Cal L$, and an {\it answer} (to ?$\bigvee \Cal H$) is a set $\Cal A \subseteq \Cal H$ such that $T\models \bigvee \Cal A$.     Our approach can be modified to compute minimal answers to such  queries   using the following result.

\p

\noindent {\bf 6.1  Theorem.}   A set $\{A_i \vert i \le r\} \subseteq \Cal H$ is contained in a minimal answer to the query ?$\bigvee \Cal H$ iff  for each $i\le r$ we may find a disjunctive stable model $M_i$ of $T$ such that $M_i\cap \{A_j \vert j \le r\}=\{A_i\}$ and $T\models \bigvee_{i\le r} A_i \vee \bigvee \bigcap_{i\le r} (\Cal H-M_i)$.

\p

As in Section 3, for stratified databases, $\{A_i \vert i \le r\} \subseteq \Cal H$ is contained in a minimal answer to the query ?$\bigvee \Cal H$ iff  for each $i\le r$ we may find a cyclic strong cover $\Cal C_i$ of $\{\lnot A_i\} \cup \{A_j \vert j\le r, j\ne i\}$ such that 
$T\models \bigvee_{i\le r} A_i \vee \bigvee (\Cal H\cap \bigcap_{i\le r} \Cal C_i^+)$.

As discussed in [Jo99, Section 5],  when $r=1$ the above theorem indicates that Theorem 1.5 does not carry over to queries of the form ?$\bigvee \Cal H$.  Specifically it enables us to show that a predicate  belongs to some minimal answer to the query ?$\bigvee \Cal H$ iff it is contained in some disjunctive stable model $M$ of $T$ for which there is no disjunctive stable model $M'$ such that $M'\cap \Cal H \subset  M\cap \Cal H$.

Computing minimal answers to a query ?$\bigvee \Cal H$ is equivalent to computing minimal answers that are contained in $\Cal H$.   This process is clearly going to be more efficient than computing all minimal answers, since our search is constrained in $\Cal H$.   
This then raises the following question.   When attempting to compute the set of all minimal answers,  might it be useful from the viewpoint of computational efficiency, to first apply a technique which generates answers $\bigvee \Cal H$, and then apply the results of the current paper to find minimal answers within $\Cal H$?

\p

\p

\noindent {\bf \S 7. \,\,   Other semantics}

\p

The perfect/disjunctive stable model semantics are of course not the only available semantics.   For both stratified and unstratified databases, we could employ the possible model semantics [Sa94] based upon inclusive (rather than exclusive) disjunction.   In this case possible models can be characterised by total supported strong covers [Jo99b], and using this characterisation, the methods of the current paper can be adapted to the possible model semantics.        For unstratified databases, possible models are ultimately defined in terms of stable models, whose possible  non-existence gives rise to the same issues as found in the current paper.     It is also worthy of note that for a certain class of stratified databases, the answers (and hence minimal answers) under the perfect model semantics coincide with those under the possible model semantics [Jo02, Corollary 7.3.7].

For unstratified databases we could alternatively employ the disjunctive 
well-founded semantics (DWFS) [Bra94].     DWFS can be constructed as the union of an increasing sequence $\emptyset = D_0\subseteq D_1\subseteq \dots$ [Bra98], where each $D_{\alpha+1}$ consists of a set $D_{\alpha+1}^+$ of disjunctions of predicates, and a set $D_{\alpha+1}^-$ of negative literals.    $D_{\alpha+1}^+$ is constructed from $D_{\alpha}^-$ which in turn is constructed from $D_{\alpha}^+$.      In [Jo01, Theorem 2.6] it is shown that  $D_{\alpha+1}^+ = \{\bigvee \Cal
P :
T\vert_{\small g} (\Cal L - \overline{D_{\alpha}^-})\models
\bigvee \Cal P\}$, and thus the computation of minimal answers {\it in each} $D_{\alpha+1}^+$ can be achieved by applying  the methods of Section 3 to the positive database $T\vert_{\small g} (\Cal L - \overline{D_{\alpha}^-})$  (using of course minimal rather than perfect models).   However, the minimality of some answer in $D_{\alpha+1}^+$ does not necessarily imply its minimality in $D_{\alpha+2}^+$ or (hence) in DWFS as a whole.    

We can overcome this problem by computing the sequence $D_0^-\subseteq D_1^-\subseteq \dots$ independently of the sets $D_{\alpha}^+$, since ([Jo01, Corollary 4.10])   $\lnot Q \in D_{\alpha+1}^-$ iff  for each   quasi-cyclic tree $\Cal T$ for
$Q$ in $T$, either $T\vert_{\small g} (\Cal L - \overline{D_{\alpha}^-})\models
\bigvee \Cal N(\Cal T)$ or $T\vert_{\small g} (\Cal L - \Cal N(\Cal T)\cup
\overline{D_{\alpha}^-})\models
\bigvee \Cal O(\Cal T)$.    Having computed DWFS$^-=\bigcup_{\alpha} D_{\alpha}^-
$, we can then compute DWFS$^+=\bigcup_{\alpha} D_{\alpha}^+ = \{\bigvee \Cal P : T\vert_g
(\Cal L - \overline{\text{DWFS}^-})\models \bigvee \Cal P\}$ [Jo01, Theorem 2.8], and in particular we can apply the techniques of Section 3 to the  positive database $T\vert_g
(\Cal L - \overline{\text{DWFS}^-})$ (again using  minimal rather than perfect models) in order 
to compute minimal answers in DWFS$^+$.  In this respect it is shown in [Jo01, Lemma 1.5]  that only predicates in $\Cal L - \overline{\text{DWFS}^-}$ can appear in a minimal answer in DWFS$^+$.

\p

\p

\noindent {\bf \S 8. \,\, The first order level}

\p

We have presented a new, and somewhat novel solution to the problem of computing minimal answers for propositional databases.   In view of the relationship with the wider issues of computational complexity we consider these results to be of interest in themselves.  

From the viewpoint of deductive databases however, ``real'' databases are of course first order (but function free).  The rules in such databases represent the set of their ground instances (this defining issues such as logical inference), but for the sake of computational efficiency we require methods that apply directly to the first order level.  In particular, grounding of the entire database, and then subsequently applying a propositional query answering method would be considered inappropriate, since the resulting ground database would often be large.

In spite of these comments, first order methods are typically developed in the first instance for the propositional level, and then lifted to the first order level.   In this sense, our results presented to-date  constitute the first step.   In this section we briefly consider the issues, difficulties and constraints that  a first order lifting encounters.   We also mention briefly an additional advantage that such a lifting would have over  the first order query processing method presented in [Jo98].

Suppose then that  $T$ is a first order database, and the set of ground instances of rules in $T$ is denoted by $gr(T)$.   A minimal answer in $T$ is simply a set of ground positive atoms that is a minimal answer in $gr(T)$.   Our basic method (i.e., without compilation) requires the computation of cyclic strong covers in $gr(T)$, so let us consider how this can be achieved (without computing $gr(T)$).

\p

\noindent {\bf 8.1  Constructing cyclic trees.}   

The definition and (top-down) construction of cyclic trees for first order (function free) databases is discussed at length in [Jo96, Jo98, Jo98a, Jo00].   Each rule within the tree is linked to the cycle above (cf., Definition 2.3(iii))  by a unifying instantiation, and a significant characteristic of cyclic trees  (for first order databases) is that these unifying instantiations have the effect of making $Pred(\Cal T) \cup \Cal O(\Cal T)\cup \Cal N(\Cal T)$ (naturally) ground.

Cyclic trees thus provide no additional complications at the first order level.  [Indeed we note as an aside that a conceptually simple approach to the generation of minimal answers is to pre-compute all cyclic trees in $T$, and (using such) transform $T$ into an equivalent positive database (see Appendix B).   Minimal answers in $T$ can be computed using strong covers (as opposed to (total) cyclic strong covers) in the transformed database (Theorem B.4), the computation of which can be achieved using the methods of [Jo97] as is discussed below.]  

\p

\noindent {\bf 8.2 Constructing cyclic strong covers.}   

Our approach to constructing cyclic strong covers at the propositional level is to employ the operator presented in Section 2.11.  Specifically given $\Cal Q$ and a rule  $C\in T$ such that conseq($C)\subseteq \Cal Q^+$ and $\Cal Q\cap (\text{antec}(C)\cup \overline{\Cal N(C)}) =  \emptyset$,  we may extend $\Cal Q$ to either \item{$\bullet$} $\Cal Q'=\Cal Q\cup \{P\}$ for some  $P\in \text{antec}(C)$, or  \item{$\bullet$}  
 $\Cal Q'=\Cal Q\cup \Cal S(\Cal T)$, where $\Cal T$ is a cyclic tree  for some  $R\in \Cal N(C)$ in $T$ such that $\Cal Q \cup \Cal S(\Cal T)$ is consistent.

As mentioned in Section 2, we can then represent the generation of cyclic strong covers as an (extended deduction) tree construction [Jo98, Jo99, Jo99a].  For example if antec($C)=\{P\}$,  $\Cal N(C) = \{R\}$ and $R$ has cyclic trees $\Cal T_1, \Cal T_2, ..., \Cal T_r$ in $T$, then the application of $C$ yields the following subtree:

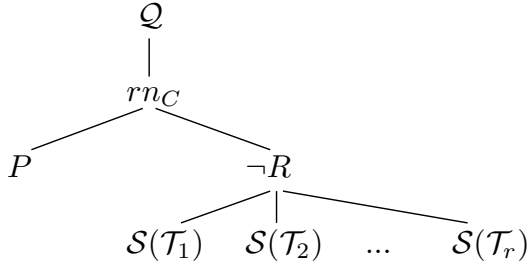
\baselineskip=24pt
$$\vbox{\settabs 10 \columns
\+&&\,\,$\Cal Q$&&&&&&&&\cr
\+&&$rn_C$\cr
\+&$P$&&$\lnot R$\cr
\+&&$\Cal S(\Cal T_1)$&$\Cal S(\Cal T_2)$&...&\hskip-10pt $\Cal S(\Cal T_r)$\cr
}$$
\baselineskip=14pt

Of course if $R$ has no cyclic tree in $T$, then we may, in effect, discard $\lnot R$ as a child of $rn_C$. 

Consider now the first order case.   As is normal in research into deductive databases we will assume that for each rule $C$, if  $x$ is a variable which appears in conseq($C$), then $x$ also appears in $\text{antec}(C) \cup \Cal N(C)$.   

Suppose for the sake of simplicity that $\Cal Q$ is ground, and that  $C$ has the form
of $P(y,z) \wedge \lnot R(y,z) \to V(y)\vee W(y)$.  We pick a substitution $\theta$ which unifies conseq($C$) with some subset of $\Cal Q$,  say $y\theta=a$.   We then compute a cyclic tree (say $\Cal T_1$) for (an instance of) $R(a,z)$, this resulting in a further instantiation $\eta$ of $z$, say $z\eta=b$.   Assuming that $\Cal Q\cap (\text{antec}(C\theta\eta) \cup \overline {\Cal N(C\theta\eta)}) = \emptyset$, we then compute the remaining cyclic trees $\Cal T_2$, ..., $\Cal T_r$ for $R(a,b)$, thus yielding the tree

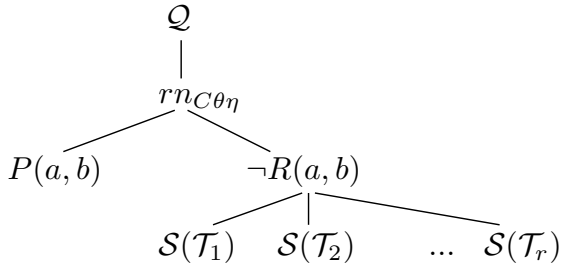
\def\k{\hskip5pt}
\def\h{\hskip10pt}
\baselineskip=24pt
$$\vbox{\settabs 10 \columns
\+&&\k\hskip-3pt$\Cal Q$&&&&&&&&\cr
\+&&$rn_{C\theta\eta}$\cr
\+&\hskip-10pt$P(a,b)$&&\hskip-10pt$\lnot R(a,b)$\cr
\+&&\h\hskip-10pt$\Cal S(\Cal T_1)$&\h\hskip-10pt$\Cal S(\Cal T_2)$&\h ... &\hskip-10pt $\Cal S(\Cal T_r)$\cr
}$$
\baselineskip=14pt

Notice that the tree is still ground, but importantly   we have  not  taken an arbitrary instance of $C$:  the instantiation of variables in conseq($C$) results from the unifier $\theta$, and the instantiation of any other variables in $\Cal N(C)$ results from the unifiers employed in the construction of the first cyclic tree $\Cal T_1$.    The issue of how to handle instances of $R(a,z)$ which have no cyclic trees  is discussed below in Section 8.3.   

Suppose instead that $C$  had the form
of $P(x,y,z) \wedge \lnot R(y,z) \to V(y)\vee W(y)$, then the resulting tree would have the form 

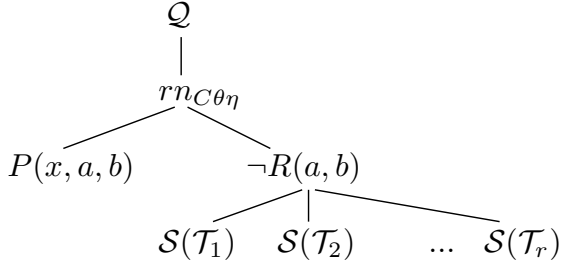
\baselineskip=24pt
$$\vbox{\settabs 10 \columns
\+&&\k\hskip-3pt$\Cal Q$&&&&&&&&\cr
\+&&$rn_{C\theta\eta}$\cr
\+&\hskip-10pt$P(x,a,b)$&&\hskip-10pt$\lnot R(a,b)$\cr
\+&&\h\hskip-10pt$\Cal S(\Cal T_1)$&\h\hskip-10pt$\Cal S(\Cal T_2)$&\h ... &\hskip-10pt $\Cal S(\Cal T_r)$\cr
}$$
\baselineskip=14pt

In this case the child nodes of $\lnot R(a,b)$ are again ground.   The left hand child of $rn_{C\theta\eta}$ contains a variable however, indicating that  this  extension of $\Cal Q$ resulting from the application of $C\theta\eta$ represents $\Cal Q\cup \{P(c,a,b) \vert c $ is a constant in $\Cal L\}$.

But now if  $C$ had the form $S(x,z) \wedge P(x,y,z) \wedge \lnot R(y,z) \to V(y)\vee W(y)$, then the application of $C\theta\eta$ results in the tree

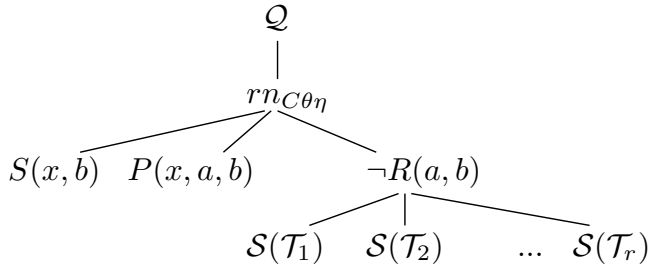
\baselineskip=24pt
$$\vbox{\settabs 10 \columns
\+&&&\k$\Cal Q$&&&&&&&&\cr
\+&&&$rn_{C\theta\eta}$\cr
\+&$S(x,b)$&$P(x,a,b)$&&$\lnot R(a,b)$\cr
\+&&&\h\hskip-10pt$\Cal S(\Cal T_1)$&\h\hskip-10pt$\Cal S(\Cal T_2)$&\h ... &\hskip-10pt $\Cal S(\Cal T_r)$\cr
}$$
\baselineskip=14pt

\noindent and again $\Cal Q\cup \{P(c,a,b) \vert c $ is a constant in $\Cal L\}$ and $\Cal Q\cup \{S(c,b) \vert c $ is a constant in $\Cal L\}$ are valid extension of $\Cal Q$.  Unfortunately, these are not the only extensions that we need to consider: for example if $\Cal L$ contains just the two constants $a$ and $b$, then $\Cal Q\cup \{S(a,b), P(b,a,b)\}$ and $\Cal Q\cup \{S(b,b), P(a,a,b)\}$ also need to be considered, and thus the branches individually do not represent the totality of all covers.  Worse still,  if $T$ contained a rule whose head had the form $S(w,b)\vee P(w,a,b)$, then such a rule would somehow have to be applied to both branches simultaneously, thus effectively destroying our view of cyclic cover construction as a tree traversal.  

A solution to this problem, presented in [Jo97, Section 4], is to insist that variables which appear in antec($C$) but not in conseq($C)\cup \Cal N(C)$ occur only in semi-definite predicates [Jo97].      Such predicates can in particular only occur in the heads of rules which are definite (i.e., contain a single atom in their head), whence this eliminates the possibility of applying the rule to two branches simultaneously.   

The problem of how to handle body variables that are not  instantiated by the unifier also occurs in the study of hyper-tableaux [Ba97], where the problem of making a blind instantiation of such variables is well known.   In [Ba97],  Baumgartner presents a solution to this problem by employing an additional {\tt Link} inference rule which effectively creates separate instantiated copies of the problematic atoms in a demand driven way, as such instantiations are required by available extension steps.    This approach is similar to that employed in [Jo97], where such variables are instantiated by so called {\it eq-rules}, again in a demand driven way.   The technical differences between the methods of [Ba97] and [Jo97] make a direct comparison difficult, but the approach presented in [Ba97] does appear to be more general, and we conjecture that it could be adapted to allow the above-mentioned restriction in terms of semi-definite predicates to be lifted.

A similar argument can be made in the case when the initial set $\Cal Q$ is non-ground (although in fact  the methods described in Section 3 could be employed without this requirement).
 
\p

\noindent {\bf 8.3  Negative subgoals with no cyclic trees.}

In the example above, the atom $\lnot R(a,z)$ becomes ground as a result of the  construction of $\Cal T_1$.   In addition, we have to also allow for instances of $R(a,z)$ that have no cyclic tree.  There are two obvious solutions to this problem:
\item{\bf (i)}    Impose a further constraint, namely that every variable appearing in $\Cal N(C)$ also appears in conseq($C$), whence the unifier $\theta$ causes $\Cal N(C)$ to become ground.
\item{\bf (ii)}  Implicitly add to the database a rule of the form $K\vee {\tt TRUE}$, for each $K\in \Cal H$.  This forces every atom to have at least one cyclic tree, but the downside of this strategy is that instances of $R(a,z)$ (which otherwise have no cyclic tree) are then chosen blindly.

\p

In the top-down query processing method presented in [Jo98] we insisted  that every variable appearing in $\Cal N(C)$ also appears in antec($C$) (this being a common assumption made of deductive databases).  This had the effect of ensuring that negative subgoals became grounded before expansion, as a result of the prior expansion of the positive subgoals arising from antec($C$).   It is unclear, although certainly worthy of investigation, whether such an assumption, together with application of the demand driven instantiation of atoms in antec($C$) (and hence in $\Cal N(C)$) as described in [Ba97],  would resolve this issue.

\p

\noindent {\bf 8.4 Answer size.} 

 In [Jo98, Jo98a], we presented a top-down query processing method applicable to first order (stratified) databases.    Suppose for example that we have a unary predicate $Q$, then the query ?$Q(x)$ represents the query ?$\bigvee\{Q(a) \vert a \text{ is a constant in  } \Cal L\}$.   An answer of the form $\bigvee_{i\le r} Q(a_i)$ is witnessed as being an answer by the absence of a cyclic strong cover, which in turn is represented by an extended deduction tree whose root is labelled with $\{Q(a_i) \vert i \le r\}$.  The difficulty in this  approach is deciding when to fix the size of the answer that we are seeking.   

Conceptually the simplest approach is to start with the goal ?$\bigvee_{i\le m} Q(x_i)$, where each $x_i$ is a variable, and $m$ is the number of constants in $\Cal L$.   As an extended deduction tree is generated, the variables $x_i$ become instantiated, and eventually yield an answer.    Of course $m$ will typically be very large, and this approach is therefore extremely inefficient (allowing massive redundancy) in the case when the answers actually being generated are much smaller than $m$.  For non-unary predicates the potential redundancy  is even worse.

The converse approach would be to first compute answers of size 1 (with initial goal ?$Q(x_1)$), then answers of size 2 (with initial goal ?$Q(x_1)\vee Q(x_2)$), and so on.   The difficulty with this approach is knowing when the process can halt, since  the computation of no new  answers of size $k$ implies nothing about the existence of minimal answers of size $k+1$.   (Indeed it is shown in [Jo96] that for propositional databases, the problem of determining whether a minimal answer of size $>k$ exists is $\Sigma_2^P$ - complete.)

The approach presented in the present paper encounters neither of these problems.

\p

\noindent {\bf 8.5  Disjunctive logic programs.}

The insistence that deductive databases are function free  is 
of course motivated by the desire that  queries should have a finite number of answers (and in the disjunctive case, that answers should themselves be finite).    
Disjunctive logic programs on the other hand allow rules with function symbols (i.e., they are full first order), whence query processing encounters the problems of first order undecidability.  This is also true of minimal answer generation and testing [Ba97a, In02].

Limitations would therefore need to be imposed on the use of function symbols to allow an extension of our methods.   Restrictions on the use of function symbols which yield a  fragment of clausal logic decidable by hyper-resolution are presented  in [Ge02], and we conjecture that similar results could be obtained in the present context.   In addition, some preliminary results on the computation of cyclic trees in infinite propositional languages  are to be found in [Jo01, Section 4].  

It is also worth noting that at the first order level, subsumption can be employed to provide alternative definitions of the notion of a minimal answer ([Ba97, In02]).   In [In02] it is shown that under such definitions, the problem of infinite answer sets can occur even in the function free case.

\p

We intend to pursue these issues related to the first order level in a sequel.

\p

\p

\noindent {\bf \S 9.   \,\,  Conclusions and open questions}

\p

We have presented a method of computing (only) minimal answers of the form $\bigvee \Cal A$ in disjunctive deductive databases (and to the best of our knowledge this is the first such method).      The method achieves this by generating (and extending) partial minimal answers, with verification being employed at each stage to ensure that a new predicate used to extend a partial answer has the required properties.     We have also discussed the problems inherent in extending the method to unstratified databases under the disjunctive stable model semantics.    The possible absence of disjunctive stable models would seem to imply the need to force  the generation of disjunctive stable models in their entirety by the addition of denial rules of the form $P\wedge \lnot P \to {\tt FALSE}$.  Compilation has been proposed as a solution to this problem, and also as a means of simplifying and improving the efficiency of the run-time computation.

The following observations and open questions are suggested by our results.

\p

\noindent {\bf 1.}     How does the method presented in the current paper compare in terms of  computational efficiency with  methods that also generate non-minimal answers?  Is it useful, again from the viewpoint of computational efficiency, to combine the two, using the latter to generate answers $\bigvee \Cal H$, and then the current method to find minimal answers within $\Cal H$?     This also raises the following question.

\p

\noindent {\bf 2.}  Let us say that a set $\{\Cal H_{\alpha} \vert \alpha \le  \beta\}$ of subsets of $\Cal L$ is a {\it minimal answer mask} iff for each  minimal answer $\Cal A$ in $T$ there is some  $\alpha\le\beta$ such that  $\Cal A\subseteq \Cal H_{\alpha}$.   

How do we determine whether a given set is a minimal answer mask?   Of course the set of all answers, and the set of all minimal answers are both minimal answer masks, and from which the computation of minimal answers is simple/trivial respectively.   Is there a method of computing minimal answer masks so that, when combined with the methods of the current paper (as suggested in 1 above), we achieve some computational gain?

\p

\noindent {\bf 3.}  What is the computational complexity of determining whether: 
\item{(i)}   there exists a non-empty cyclic strong cover;  
\item{(ii)}  a given literal belongs to some maximal cyclic strong cover;  
\item{(iii)}  every cyclic strong cover can be extended to a disjunctive stable model;
\item{(iv)}   a given literal belongs to some cyclic strong cover $\Cal C$ for which there is no cyclic strong cover $\Cal D$ such that $\Cal D^+\cup \Cal D^-\supset \Cal C^+\cup \Cal C^-$;
\item{(v)}  a given set is a minimal answer mask;
\item{(vi)}  given a formula $\Phi$, there is a subset $T^*\subseteq T$ such that $T^*$ has a disjunctive stable model, and $T^*\models \Phi$; 
\item{(vii)}  given a formula $\Phi$, there is a subset $T^*\subseteq T$ such that $T^*$ has a disjunctive stable model $M$ with $M\models \Phi$.

The question given in (iv) is perhaps somewhat artificial, but it is of interest since a similar problem is shown in [Ei98] to be $\Sigma_3^P$ - complete.   Questions (vi)/(vii) ask whether there is a ``consistent'' subset of $T$ with which $\Phi$ can be inferred/is consistent.  The property given in (vi) has application in argumentation-based dialogues in multi-agent systems [Pa02].

\p

\noindent {\bf 4.}      In [Pr91, Pr91a], Przymusinski introduced the notion of a disjunctive stationary model, these being consistent sets of literals $I$ such that $I^+$ is a minimal model of $T\vert_g (\Cal L - I^-)$ and $\Cal L - I^-$ is a minimal model of $T\vert_g I^+$.  Total disjunctive stationary models clearly coincide with disjunctive stable models, and thus disjunctive stationary models can (also) be regarded as partial disjunctive stable models.  

For non-disjunctive databases, stationary models are related to (supported) strong covers [Jo02, Section 4.2], but in the disjunctive case it is doubtful that any such relationship exists with cyclic strong covers.   For example if $T=\{C\vee E, B\to D, B\to A, C\wedge \lnot E \to A\vee B, A\wedge C \wedge \lnot D \to B\vee E\}$, then $\{A,C,\lnot E\}$ is a disjunctive stationary model, yet there is no cyclic strong cover containing $\lnot A$.

Similarly the analogue of Theorem 2.9 does not hold for disjunctive stationary models.   For example 
 if $T=\{C\vee E,  D \to A,  \lnot D \to B, \lnot B \to D, C\wedge \lnot E \to A \vee B\}$, then again $I=\{A,C,\lnot E\}$ is a disjunctive stationary model, and indeed $\overline {I} = \{\lnot A, \lnot C, E\}$ is a strong cover.   $T_{\overline{I}} = \{\lnot D\to B, \lnot B\to D\}$ (Definition 2.8) has disjunctive  stable models $\{B\}$ and $\{D\}$, but  $\{A,B,C\}$ is not a disjunctive stable model of $T$.

Ultimately, the difference between disjunctive stationary models and cyclic strong covers lies in the motivation for their definition.   Cyclicness insists that 
$\Cal C^-$ is contained with {\it any} model of $T\vert_g (\Cal L - \Cal C^+)\wedge \lnot \bigvee \Cal C^+$, whereas stationarity insists that $I^+$ is one (of possibly many) minimal models of $T\vert_g (\Cal L - I^-)$.

\p

\noindent {\bf 5.}    As indicated in Section 7, DWFS can be characterised using the  variant notion of quasi-cyclic trees.    We note however that again there appears to be no direct relationship between DWFS and cyclic strong covers.   For example if $T=\{A\vee D, B\vee E, C\vee F, \lnot A \to B, \lnot B \to C, \lnot C \to A, \lnot D \to E, \lnot E \to F, \lnot F \to D\}$, then there is no non-empty cyclic strong cover, yet DWFS = $\{A\vee D, B\vee E, C\vee F\}$.  

As an aside we recall that in [Jo01, Section 2] we showed that if $I$ is a disjunctive stationary model, then $I^+\models $DWFS$^+$ and $I^-\supseteq \overline{\text{DWFS}^-}$, which in turn suggested the following question:   ``Given a minimal  model $M$ of DWFS$^+$, is  it the case that $M\cup \text{DWFS}^-$ can always be extended to a disjunctive stationary model?''   The above database, having no disjunctive stationary models, shows the answer to this question to be negative.  

\p

\noindent {\bf 6.}   If $T$ is stratified, and $\Cal S = ((A_i, \Cal C_i) \vert i \le r)$ is a verified cyclic state, is it the case that we can construct a reduced database $T[\Cal S]$  such that the complete cyclic states in $T[\Cal S]$ characterise the complete extensions of $\Cal S$  in $T$?

\p

\noindent {\bf 7.}  Suppose that $T$ is unstratified, and that $\Cal S = ((A_i, \Cal C_i) \vert i \le r)$ is a verified cyclic state.   Suppose that some  $\Cal C_i$ cannot be extended to a total cyclic strong cover, then (by the remarks following Theorem 2.7) $T\models \bigvee \Cal C_i^+$, where $\Cal C_i^+\supseteq \{A_j \vert j \le r, j\ne i\}$.   What can we then infer (if anything)  about the minimal answers  related to $\Cal S$?

The only obvious (and weak) statement we can make, is that each $A_i$ is contained in some minimal answer in $T/\Cal C_i$ (by Theorem 2.7).

\p 

\noindent{\bf 8.}  What is the relationship (if any) between the methods of the current paper, and other $\Sigma_2^P$ - complete problems, for example those relating to argumentation-based dialogues in multi-agent systems [Pa02].

\p

\noindent {\bf 9.}  Is there a weaker property than stratification which guarantees that every cyclic strong cover may be extended to a disjunctive stable model?

\p

\noindent {\bf 10.}   Our characterisations (e.g., Theorem 3.3) of minimal answers can easily be extended to the case when we consider answers to be disjunctions of literals (as opposed to disjunctions of predicates).  We have not however detailed further the extension of our results to this case.

\p

\p

\noindent {\bf References}

\item{[Ba97]}  P.  Baumgartner,  Hyper-tableaux - The next generation, Technical Report 32/97, Institut f\"ur Informatik,  Universit\"at Koblenz-Landau (1997). 
\item{[Ba97a]}  P. Baumgartner, U. Furbach and F. Stolzenburg, Computing answers with model elimination, Artificial Intelligence, vol. 90 (1997), 135-176. \item{[Bra94]} S. Brass and J. Dix, A disjunctive semantics based
upon unfolding and bottom-up evaluation, in : B. Wolfinger, (ed.),
Innovationen bei Rechen- und Kommunikationssystemen (IFIP-
Congress,
Workshop FG2: Disjunctive Logic Programming and Disjunctive
Databases), (Springer, 1994), 83-91.
\item{[Bra95]}   S. Brass and J. Dix, A general approach to bottom-up computation of disjunctive semantics, in J. Dix, L. Pereira, T. Przymusinski (eds.): Non-Monotonic Extensions of Logic Programming, Springer Lecture Notes in Artificial Intelligence, vol. 927 (Springer, Berlin, 1995), 127-155.
\item{[Bra98]}  S. Brass and J. Dix, Characterisations of the
disjunctive well-founded semantics: confluent calculi and
iterated GCWA, J. Automated Reasoning, vol. 20 (1998), 143-165.
\item{[Gr86]} J. Grant and J. Minker, Answering queries in
indefinite databases
and the null value problem, Advances in Computing Research,
vol. 3 (1986),
247-267.
\item{[Ei93]}  T. Eiter and G. Gottlob, Propositional
circumscription and extended closed world reasoning are $\prod
_2^P$ - complete, Theoretical Computer Science, vol. 114 (1993)
231 -  245.
\item{[Ei98]}  T.  Eiter, N. Leone and D. Sacc\`a,  Expressive power and complexity of partial models of disjunctive deductive databases,  Theoretical Computer Science, 206 (1998), 181-218.
\item{[Fe95]}  J. Fern\'andez, J. Minker and A. Yahya, Computing perfect and stable models using ordered model trees,  Computational Intelligence, vol. 11 (1995), 89-112.
\item{[Fe95a]}  J. Fern\'andez and J. Minker, Bottom-up computation of perfect models for disjunctive theories,  Journal of Logic Programming, vol. 25 (1995),  33-51.
\item{[He88]} L.J. Henschen and H. Park, Compiling the GCWA in
indefinite databases, in J. Minker, ed., Foundations of Deductive
Databases, pp 395-438 (Morgan Kaufmann, Washington), 1988. 
\item{[Ge88]}  M. Gelfond and V. Lifschitz, The stable model
semantics for logic programming, in: R. Kowalski and K. Bowen
(eds.), Proc. 5th International Conference on Logic Progamming,
Seattle
(1988), 1070-1080.
\item{[Ge02]}  L. Georgieva, U. Hustadt and R.A. Schmidt,  A new clausal class decidable by hyperresolution,  University of Manchester Department of Computer Science technical report CSPP-18.  Available May, 2003:  http://www.cs.man.ac.uk/$\sim$schmidt/.
\item{[In02]}  K. Inoue and K. Iwanuma, Minimal Answer Computation and SOL, in:  S. Greco and N. Leone (eds.), Logics in Artificial Intelligence: Proceedings of the Eighth European Conference, (Springer Lecture Notes in Artificial Intelligence, vol. 2424, 2002), 245-257. 
\item{[Jo96]} C.A. Johnson, On computing minimal and perfect
model membership, Data and Knowledge Engineering, vol. 18 (1996),
225-276. 
\item{[Jo97]}  C.A.  Johnson, Deduction trees and the view update problem in indefinite deductive databases, J. Automated Reasoning, vol. 19 (1997), 31-85. 
\item{[Jo98]} C.A. Johnson,  Top down query processing in
indefinite
deductive databases, Data and Knowledge Engineering, vol. 26
(1998), 1-36.
\item{[Jo98a]}   C.A. Johnson, Extended deduction trees and query processing, Keele University technical report TR98-07. Available May, 2003: http://www.tech.plym.ac.uk/\newline soc/staff/chrisjohnson/tr95.html 
\item{[Jo99]}  C.A. Johnson, On cyclic covers and perfect
models,  Data and Knowledge Engineering, vol. 31 (1999),
25-65.
\item{[Jo99a]}  C.A. Johnson, Processing deductive databases under the disjunctive stable model semantics, Fundamenta Informaticae, vol. 40 (1999), 31-51. 
\item{[Jo00]}   C.A. Johnson, Top-down query processing in first order deductive databases under the DWFS, in Z. Ras and S. Ohsuga (eds), Foundations of Intelligent Systems, 12th International Symposium on Methodologies for Intelligent Systems, Charlotte, USA (Springer, 2000), 377-388.
\item{[Jo01]}  C.A. Johnson,  On the computation of the disjunctive well-founded semantics,  J. Automated Reasoning, vol.  26 (2001),  333-356.
\item{[Jo02]}    C.A. Johnson, Processing indefinite deductive databases under the possible model semantics, Fundamenta Informaticae, vol. 49 (2002), 325-347. 
\item{[Jo03]}  C.A. Johnson,  Query compilation under the disjunctive well-founded semantics, pre-print.  Available May, 2003:  
http://www.tech.plym.ac.uk/soc/staff/\newline chrisjohnson/compilation/compilation.pdf 
\def\paper{
\item{[Jo03a]}  C.A. Johnson,  On the computation of only minimal answers, Extended technical report.  Available May, 2003:  
http://www.tech.plym.ac.uk/soc/staff/\newline chrisjohnson/minimal/extr.pdf }

\x
\item{[Lb92]}  J. Lobo, J. Minker and A. Rajasekar, Foundations
of Disjunctive Logic Programming, (MIT Press, Cambridge,
Massachusetts, 1992).
\item{[Mi82]}  J. Minker,  On indefinite databases and the closed world assumption, in D.W. Loveland (ed.): Proceedings of the 6th Conference on Automated Deduction,  Springer Lecture Notes in Computer Science, vol. 138 (Spring, Berlin, 1982), 292-308.
\item{[Pa02]}  S. Parsons, M. Wooldridge and L. Amgoud,  An analysis of formal inter-agent dialogues, in:  P. McBurney and M. 
Wooldridge (eds.),  Proceedings of the Fifth UK Workshop in Multi-Agent Systems (Liverpool, 2002).
\item{[Pr88]} T. Przymusinski, On the declarative semantics of
deductive databases and programs, in: J. Minker (Ed\.),
Foundations of Deductive Databases and Logic Programming, (Morgan
Kauffman, Washington, 1988).
\item{[Pr89]}  T. Przymusinski, On the declarative and
procedural semantics of logic programs, J. Automated Reasoning,
vol. 5 (1989), 167-205.   
\item{[Pr91]}  T. Przymusinski, Stable semantics for
disjunctive databases, New Generation Computing, vol. 9 (1991),
401-424.
\item{[Pr91a]}  T.  Przymusinski,   Stationary semantics for normal and disjunctive logic programs,  in:  C.  Delobel, M. Kifer and Y. Masunaga (eds.),   Proceedings of the Second International Conference on Deductive and Object-Oriented Databases DOOD'91, Munich (Springer Verlag, 1991), 85-107.
\item{[Ra89]} A. Rajasekar, Semantics for Disjunctive Logic Programs, PhD thesis, University of Maryland (1989).
\item{[Sa94]}  C. Sakama and K. Inoue, An alternative approach
to the semantics of disjunctive logic programs and deductive
databases, J. Automated Reasoning, vol. 13 (1994), 145-172.
\item{[St77]} L. Stockmeyer, The polynomial-time hierarchy, 
Theoretical Computer Science, vol. 3 (1977), 1-22.
\item{[Wa86]} K. Wagner and G. Wechsung, Computational Complexity
(Reidel, 1986).
\item{[Wr77]} C. Wrathall, Complete sets and the polynomial-time
hierarchy, Theoretical Computer Science, vol. 3 (1977), 23-33.
\item{[Ya94]} A. Yahya, J. Fern\'andez and J. Minker, Ordered
model trees : A normal form for disjunctive deductive databases,
J. Automated Reasoning, vol. 13 (1994), 117-143.
\item{[Ya96]}  A. Yahya, A goal-driven approach to efficient query processing in disjunctive databases, Technical Report PMS-FB-1996-12, Institut f\"ur Informatik, Ludwig Maximilians Universit\"at, M\"unchen, Germany.  
\item{[Ya02]}  A. Yahya,  Duality for efficient query processing in disjunctive deductive databases,  J. Automated Reasoning, vol. 28 (2002). 

\p

\p

\noindent {\bf Appendix A:  Worked examples}

Let INT$(\Cal L) = \{Q_1, Q_2, Q_3, Q_4\}$ and INT($T$) consist of the following rules

$$\vbox{\settabs 20 \columns
\+1. \,\,   $Q_2\wedge Q_3 \wedge \lnot R_1 \to Q_1\vee Q_4$&&&&&&&&2. \,\,  $Q_1\wedge \lnot R_2\to Q_2$&&&&&&3. \,\, $S_2 \wedge \lnot R_3 \to Q_3$&&&&\cr
\+4. \,\,  $S_3 \to Q_1 \vee Q_2 \vee Q_4$&&&&&&&&5. \,\,  $S_1 \to Q_2\vee Q_3$&&&&&&6. \,\, $R_1\to Q_2$\cr
}$$
then the partial cyclic trees (cf., Section 5) in INT($T$) are as follows.  
$$\vbox{\settabs 25 \columns
\+&$Q_1$&&&&$Q_1$&&&$Q_2$&&&$Q_2$&&&$Q_2$&&&$Q_3$&&&$Q_3$&&&$Q_4$&\cr
\+&$rn_1$&&&&$rn_4$&&&$rn_6$&&&$rn_5$&&&$rn_4$&&&$rn_3$&&&$rn_5$&&&$rn_4$\cr
\+$Q_2$&&$Q_3$&&&$S_3$&&&$R_1$&&&$S_1$&&&$S_3$&&&$S_2$&&&$S_1$&&&$S_3$\cr
\+$rn_2$&&$rn_3$&&&&&&&&&&&\cr
\+$Q_1$&&$S_2$&&&$\Cal T_2$&&&$\Cal T_3$&&&$\Cal T_4$&&&$\Cal T_5$&&&$\Cal T_6$&&&$\Cal T_7$&&&$\Cal T_8$\cr
\+$rn_4$&&&&&&&&&&&&&&&&&&&&\cr
\+$S_3$&&&&&&&&&&&&&&&&&&&&\cr
\+&$\Cal T_1$&&&\cr
}$$

There is one further tree (for $Q_2$) which yields the same $\Cal S(\Cal T)$ set as $\Cal T_1$, whence we omit it.
Below we present, for each tree, the set $\Cal S(\Cal T)$.

\item{\bf 1.}  $\Cal T_1$ : $\{\lnot Q_1, \lnot Q_2, \lnot Q_3, Q_4, R_1, R_2, R_3, \lnot S_2, \lnot S_3\}$ 
\item{\bf 2.}  $\Cal T_2$ : $\{\lnot Q_1, \lnot S_3, Q_2, Q_4\}$
\item{\bf 3.}  $\Cal T_3$ : $\{\lnot Q_2, \lnot R_1\}$
\item{\bf 4.}  $\Cal T_4$ :  $\{\lnot Q_2, \lnot S_1, Q_3\}$
\item{\bf 5.}   $\Cal T_5$ :  $\{\lnot Q_2, \lnot S_3, Q_1, Q_4\}$
\item{\bf 6.}  $\Cal T_6$ : $\{\lnot Q_3, \lnot S_2, R_3\}$
\item{\bf 7.}  $\Cal T_7$ : $\{\lnot Q_3, \lnot S_1, Q_2\}$
\item{\bf 8.}   $\Cal T_8$ : $\{\lnot Q_4, \lnot S_3, Q_1, Q_2\}$.

By Theorem 2.6(c) (and the characterisation of weakly cyclic covers in terms of cyclic strong covers), if $\Cal C$ is a weakly cyclic cover containing $\lnot Q_i$, then there must be some partial tree $\Cal T$ for $P$ such that $\Cal S(\Cal T)\subseteq \Cal C$.   (By the above remark,  $\Cal T_1$ must also count as a tree for $Q_2$.)

As indicated in Sections 2, we can depict the generation of weakly cyclic covers using (the analogues) of the operations presented in Section 2.11:  in particular, negative atoms are expanded with the relevant $\Cal S(\Cal T)$ sets.  For the sake of brevity in this example, we first compute the minimal strong covers of the above $\Cal S(\Cal T)$ sets in INT($T$):

\item{\bf 1.}  $\Cal C_1 = \{\lnot Q_1, \lnot Q_2, \lnot Q_3, Q_4, R_1, R_2, R_3, \lnot S_2, \lnot S_3\}$ 
\item{\bf 2.}  $\Cal C_2 = \{\lnot Q_1, \lnot S_3, Q_2, Q_4, \lnot R_2, R_1\}$
\item{\bf 3.}  $\Cal C_3 = \{\lnot Q_2, \lnot R_1\}$
\item{\bf 4.}  $\Cal C_{4a} = \{\lnot Q_2, \lnot S_1, Q_3, S_2\}$  and   \newline $\Cal C_{4b} = \{\lnot Q_2, \lnot S_1, Q_3, \lnot R_3\}$
\item{\bf 5.}   $\Cal C_{5a} = \{\lnot Q_2, \lnot S_3, Q_1, Q_4, Q_3, S_2\}$, \newline $\Cal C_{5b} = \{\lnot Q_2, \lnot S_3, Q_1, Q_4, Q_3, \lnot R_3\}$, \newline $\Cal C_{5c} = \{\lnot Q_2, \lnot S_3, Q_1,$ $ Q_4, \lnot R_1\}$
\item{\bf 6.}  $\Cal C_6 = \{\lnot Q_3, \lnot S_2, R_3\}$
\item{\bf 7.}  $\Cal C_{7a} = \{\lnot Q_3, \lnot S_1, Q_2, Q_1, R_1\}$, \newline   $\Cal C_{7b} = \{\lnot Q_3, \lnot S_1, Q_2, \lnot R_2, R_1\}$
\item{\bf 8.}   $\Cal C_8 = \{\lnot Q_4, \lnot S_3, Q_1, Q_2, R_1\}$.

Thus if $\Cal C$ is a weakly cyclic cover containing $\lnot Q_i$, then it must contain one of (i)  $\Cal C_1, \Cal C_2$, when $i=1$; (ii)  $\Cal C_1, \Cal C_3, \Cal C_{4a}, \Cal C_{4b}, \Cal C_{5a}, \Cal C_{5b}, \Cal C_{5c}$, when $i=2$;  (iii)  $\Cal C_6, \Cal C_{7a}, \Cal C_{7b}$ when $i=3$; and (iv) $\Cal C_8$ when $i=4$.

We represent the int-total weakly cyclic covers as the branches through the following tree. 

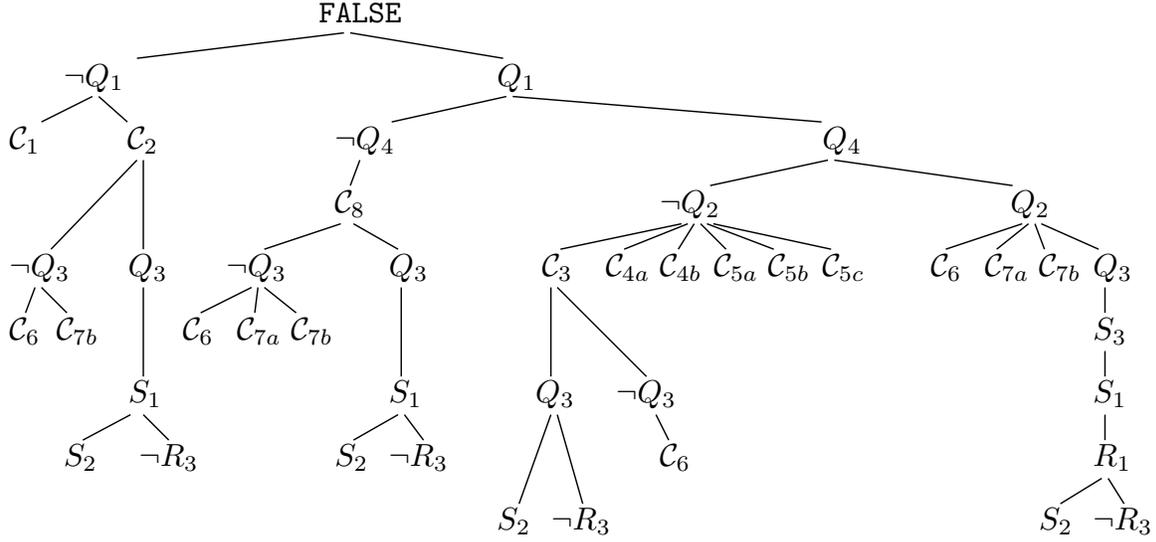
\baselineskip=20pt

$$\vbox{\settabs 22\columns
\+&&&&&&\hskip-5pt{\tt FALSE}&&&&&&&&&&&&&&&\cr
\+&$\lnot Q_1$&&&&&&&&$Q_1$&&&&&&&&&&&\cr
\+$\Cal C_1$&&\hskip3pt$\Cal C_2$&&&&$\lnot Q_4$&&&&&&&&&$Q_4$&&&&&\cr
\+&&&&&&$\Cal C_8$&&&&&&$\lnot Q_2$&&&&&&\hskip8pt$Q_2$&\cr
\+$\lnot Q_3$&&\,\,$Q_3$&&$\lnot Q_3$&&&$Q_3$&&&\hskip-3pt$\Cal C_3$&$\Cal C_{4a}$&$\Cal C_{4b}$&$\Cal C_{5a}$&$\Cal C_{5b}$&$\Cal C_{5c}$&&$\Cal C_6$&$\Cal C_{7a}$&$\Cal C_{7b}$&$Q_3$&\cr
\+$\Cal C_6$\,\,\,$\Cal C_{7b}$&&&\,\,$\Cal C_6$&\,\,$\Cal C_{7a}$&\,\,$\Cal C_{7b}$&&&&&&&&&&&&&&&$S_3$\cr
\+&&\,\,$S_1$&&&&&$S_1$&&&\hskip-5pt$Q_3$&\,\,$\lnot Q_3$&&&&&&&&&$S_1$\cr
\+&$S_2$&\,\,\,\,$\lnot R_3$&&&&$S_2$&$\lnot R_3$&&&&&$\Cal C_6$&&&&&&&&$R_1$\cr
\+&&&&&&&&&$S_2$&$\lnot R_3$&&&&&&&&&$S_2$&$\lnot R_3$\cr
\+&&&&&&&&&&&&&&&&&&&&\cr
}$$

\baselineskip=14pt

Note that the path to  $\Cal C_{5c}$ is not int-total, but has not been expanded since the set of literals  along the path  contains the set of literals along the path to the sibling $\Cal C_3$, the extensions of which have already been generated.  Notice also that in expanding $\lnot Q_2$ we have not employed $\Cal C_1$, since it would generate an inconsistent branch.

Although such trees allow us to  represent weakly cyclic covers, it is unclear whether they provide a means of {\it choosing} cyclic covers during the computation of minimal answers.

Below we enumerate the int-total weakly cyclic covers identified.   The reader will note that in some of these sets we have (for the sake of readability) left duplicates in.  The notes on the right indicate the branch of the above tree.

\def\hhfill{\hfill}

\noindent{\bf Case A.}  $\lnot Q_1\in \Cal C$

\item{1.}  $\{\lnot Q_1, \lnot Q_2, \lnot Q_3, Q_4, R_1, R_2, R_3, \lnot S_2, \lnot S_3\}$ \hhfill $\Cal C_1$
\item{2.}  $\{\lnot Q_1, \lnot S_3, Q_2, Q_4, \lnot R_2, R_1, \lnot Q_3, \lnot S_2, R_3\}$ \hhfill $\Cal C_2, \Cal C_6$
\item{3.}$\{\lnot Q_1, \lnot S_3, Q_2, Q_4, \lnot R_2, R_1, \lnot Q_3, \lnot S_1, Q_2, \lnot R_2, R_1\}$ \hhfill $\Cal C_2, \Cal C_{7b}$
\item{4.} $\{\lnot Q_1, \lnot S_3, Q_2, Q_4, \lnot R_2, R_1, Q_3, S_1, S_2\}$ \hhfill  $\Cal C_2, Q_3\in \Cal C$
\item{5.}  
$\{\lnot Q_1, \lnot S_3, Q_2, Q_4, \lnot R_2, R_1, Q_3, S_1, \lnot R_3\}$
\hhfill $\Cal C_2, Q_3\in \Cal C$

\noindent {\bf Case B.}  $Q_1\in \Cal C$ and $\lnot Q_4\in \Cal C$

\item{6.}  $\{Q_1, \lnot Q_4, \lnot S_3, Q_1, Q_2, R_1, \lnot Q_3, \lnot S_2, R_3\}$
\hhfill $\Cal C_8, \Cal C_6$
\item{7.}   $\{Q_1, \lnot Q_4, \lnot S_3, Q_1, Q_2, R_1,  \lnot Q_3, \lnot S_1, Q_2, Q_1, R_1\}$ \hhfill  $\Cal C_8, \Cal C_{7a}$

\item{8.}
 $\{Q_1, \lnot Q_4, \lnot S_3, Q_1, Q_2, R_1, \lnot Q_3, \lnot S_1, Q_2, \lnot R_2, R_1\}$  \hhfill  $\Cal C_8,\Cal C_{7b}$

\item{9.}  $\{Q_1, \lnot Q_4, \lnot S_3, Q_1, Q_2, R_1, Q_3, S_1, S_2\}$  \hhfill $\Cal C_8, Q_3\in \Cal C$

\item{10.}   $\{Q_1, \lnot Q_4, \lnot S_3, Q_1, Q_2, R_1, Q_3, S_1, \lnot R_3\}$ 
\hhfill $\Cal C_8, Q_3\in \Cal C$

\noindent {\bf Case C.}    $Q_1\in \Cal C$,  $Q_4\in \Cal C$ and $\lnot Q_2\in \Cal C$

\item{11.}  $\{Q_1, Q_4, \lnot Q_2, \lnot R_1, Q_3,S_2\}$  \hhfill $\Cal C_3, Q_3\in \Cal C$

\item{12.}   $\{ Q_1, Q_4, \lnot Q_2, \lnot R_1, Q_3,\lnot R_3\}$ \hhfill $\Cal C_3, Q_3\in \Cal C$

\item{13.}   $\{ Q_1, Q_4, \lnot Q_2, \lnot R_1,  \lnot Q_3, \lnot S_2, R_3\}$
\hhfill  $\Cal C_3, \Cal C_6$
\item{14.}   $\{ Q_1, Q_4, \lnot Q_2, \lnot S_1, Q_3, S_2\}$  \hhfill  $\Cal C_{4a}$

\item{15.}  $\{ Q_1, Q_4, \lnot Q_2, \lnot S_1, Q_3, \lnot R_3\}$  \hhfill $\Cal C_{4b}$

\item{16.}   $\{ Q_1, Q_4, \lnot Q_2, \lnot S_3, Q_1, Q_4, Q_3, S_2\}$  \hhfill $\Cal C_{5a}$
 
\item{17.}    $\{ Q_1, Q_4, \lnot Q_2, \lnot S_3, Q_1, Q_4, Q_3, \lnot R_3\}$\hhfill $\Cal C_{5b}$

\item{18.}   $\{ Q_1, Q_4, \lnot Q_2, \lnot S_3, Q_1, Q_4, \lnot R_1\}$\hhfill $\Cal C_{5c}$

\noindent {\bf Case D.}   $Q_1\in \Cal C$, $Q_4\in \Cal C$ and $Q_2\in \Cal C$

\item{19.}  $\{ Q_1, Q_4, Q_2, \lnot Q_3, \lnot S_2, R_3, S_3, R_1\}$\hhfill $\Cal C_{6}$

\item{20.}  $\{ Q_1, Q_4, Q_2, \lnot Q_3, \lnot S_1, Q_2, Q_1, R_1, S_3\}$ \hhfill $\Cal C_{7a}$

\item{21.}  $\{ Q_1, Q_4, Q_2, \lnot Q_3, \lnot S_1, Q_2, \lnot R_2, R_1, S_3\}$\hhfill $\Cal C_{7b}$

\item{22.}  $\{ Q_1, Q_4, Q_2, Q_3, S_3, S_1, R_1, S_2\}$ \hhfill $Q_3\in \Cal C$

\item{23.}   $\{ Q_1, Q_4, Q_2, Q_3, S_3, S_1,R_1,  \lnot R_3\}$\hhfill $Q_3\in \Cal C$

\p

\noindent {\bf A.1 Example.}   Suppose now that EXT($T$) contains a single rule $R_1\vee S_3$.  There are then precisely 3 minimal int-total cyclic strong covers:

$\Cal D_1 = \{Q_1, \lnot Q_4, \lnot S_3, R_1, Q_2, Q_3, S_1, S_2\}$  \hhfill from 9

$\Cal D_2 = \{Q_1, Q_4, \lnot Q_2, \lnot R_1, S_3, Q_3,S_2\}$  \hhfill from 11

$\Cal D_3 = \{Q_1, Q_4, \lnot Q_2, \lnot S_3, R_1, Q_3, S_2\}$  \hhfill from 16

\noindent [Note that $\Cal D = \{Q_1, Q_4, Q_2, Q_3, S_3, S_1, R_1, S_2\}$ is a completion of (22), but it is not a strong cover since EXT($T)\models \bigvee \Cal D_{ext}^+$.]

Note that $Q_1$ and $Q_3$ do not therefore belong to any minimal answer.     $((Q_2, \Cal D_2))$ is a (verified) cyclic state of length 1, thus let us consider its extension using the approach detailed in  Section 5.7.3.     We pick an int-total cyclic strong cover $\Cal C$ of $\{Q_2\}$, the only choice being $\Cal C = \Cal D_1$.

Suppose first that we intend to (try to) extend using a predicate $Q$ in INT($\Cal L$), then we require that $Q\in \Cal D_1^-\cap \Cal D_2^+$, and (not surprisingly) the only option is $Q_4$.  We may then extend the cyclic state to $((Q_2, \Cal D_2), (Q_4, \Cal D_1))$.   There are no int-total cyclic strong covers of $\{Q_2, Q_4\}$, whence $Q_2\vee Q_4$ is a minimal answer (and clearly is the only minimal answer contained in INT($\Cal L$)).

Suppose that we now try to extend $((Q_2, \Cal D_2))$ with a predicate $E$ in EXT($\Cal L$).   Section 5.7.3 demands that $E\not\in \Cal D_1^+\cup \Cal D_2^-$, and also that there exists a rule $\bigvee \Cal E\in \text{EXT}(T)$ such that $E\in \Cal E$ and EXT($T)\not\models \bigvee (\Cal E-\{E\})\vee \bigvee \Cal (\Cal D_1)_{ext}^+$.     $S_3$ satisfies the conditions of Definition 5.7.3, with $(\Cal E-\{E\})\cup \Cal D_1=\Cal D_1$, and $((Q_2, \Cal D_2), (S_3, \Cal D_1))$ is an immediate extension of $((Q_2, \Cal D_2))$.   $\{Q_2, S_3\}$ has no int-total cyclic strong cover, whence $Q_2\vee S_3$ is a minimal answer.

\p

\noindent {\bf A.2  Example.}   Let  EXT($T) = \{S_3\vee R_2, R_1\vee R_2, S_1\vee S_2\vee R_3\}$.  

Suppose that we start with the cyclic strong cover $\Cal F_1=\{Q_1, Q_4, Q_2, \lnot Q_3, \lnot S_2, S_1,$ $ R_3, S_3, R_1\}$ of $\{\lnot Q_3\}$ (by completing 19).   We then look for an int-total cyclic strong cover of $\{Q_3\}$, say $\Cal F_2=\{Q_1, Q_4, \lnot Q_2, \lnot R_1, R_2, Q_3,S_2\}$ (by completing 11).    

If we wish to extend using a predicate in INT($\Cal L)$, then we pick such a predicate in $\Cal F_2^-\cap \Cal F_1^+$, the only option being $Q_2$.  
$\Cal S_1=((Q_3, \Cal F_1), (Q_2, \Cal F_2))$ is not verified, since (for example) $\Cal G_1 = \{ Q_1, Q_4, Q_2, Q_3, S_3, S_1, R_1, \lnot R_3, S_2\}$ (from 23) is an int-total cyclic strong cover of $\{Q_2, Q_3\}\cup (\Cal F_1^+\cap \Cal F_2^+) = \{Q_2, Q_3, Q_1, Q_4\}$.   If $\Cal S_1$ is verifiable, then (by Theorem 5.6.1) there is a rule $\bigvee \Cal E\in \text{EXT}(T)$ such that $(\Cal E-\Cal G_1)\cap (\Cal F_1^-\cup \Cal F_2^-)=\emptyset$,  $\Cal E\cap (\Cal G_1 - \Cal F_i)\ne\emptyset$ and EXT($T)\not\models \bigvee (\Cal F_i^+\cup (\Cal E - \Cal G_1))_{ext}$.    $S_1\vee S_2\vee R_3$ is such a rule, and therefore $((Q_3, \Cal F_1\cup \{R_3\}), (Q_2, \Cal F_2\cup \{R_3\}))$ is an immediate (and verified) extension of $\Cal S_1$, which can then be extended using the methods of Section 5.7.

\p

\noindent {\bf A.3 Example.}  Suppose again that  EXT($T) = \{S_3\vee R_2, R_1\vee R_2, S_1\vee S_2\vee R_3\}$ and we start with the cyclic strong cover $\Cal F_1=\{Q_1, Q_4, Q_2, \lnot Q_3, \lnot S_2, S_1,$ $ R_3, S_3, R_1\}$ of $\{\lnot Q_3\}$ (by completing 19).   We then look for a cyclic strong cover of $\{Q_3\}$, say  $\Cal G_1 = \{ Q_1, Q_4, Q_2, Q_3, S_3, S_1, R_1, \lnot R_3, S_2\}$ (from 23).   

If we attempt to  extend $((Q_3, \Cal F_1))$ using a predicate in EXT($\Cal L)$, then    $R_3$ satisfies the conditions of Section 5.7.3, and $\Cal S_2=((Q_3, \Cal F_1), (R_3, \Cal G_1))$ is an immediate  extension of  $((Q_3, \Cal F_1))$.

It is easy to check that $\{Q_3, R_3\}\cup (\Cal F_1^+\cap \Cal G_1^+) = \{Q_3, R_3, Q_1, Q_2, Q_4, S_1, S_3, R_1\}$ has no int-total cyclic strong cover, and thus $\Cal S_2$ is verified.

Again,  $\{Q_3, R_3\}$ has an int-total cyclic strong cover, for example 
$\Cal G_2=\{Q_1, Q_4,$ $ \lnot Q_2, \lnot S_1, S_2, R_3, Q_3\}$ (from 14).     We look for a predicate in EXT($\Cal L) - (\Cal G_2^+\cup \Cal F_1^-\cup \Cal G_1^-) =    \{S_1, S_3, R_1,$ $ R_2\}$ satisfying the conditions of Section 5.7.3.    
$S_1$ satisfies these  conditions, and  $((Q_3, \Cal F_1), (R_3, \Cal G_1), (S_1, \Cal G_2))$ is an immediate extension of $\Cal S_2$.   It is easy to check that $\{Q_3, R_3, S_1\}$ has no int-total cyclic strong cover, whence is a minimal answer.

\p

\p

\noindent {\bf Appendix B:  Transformation to a positive database.}

\p

In [Fe95] it is shown that the disjunctive stable models of $T$ can be computed from the perfect models of the {\it stratified evidential transformation} of $T$.    In [Jo99]  we employed cyclic trees (and variants there-of) to present a number of forms of pre-processing for stratified databases.   In this section we extend one of these forms to the disjunctive stable model semantics.    Specifically we show that the disjunctive stable models (and minimal answers) of $T$ can be computed (in a one-to-one fashion) from the minimal models (and minimal answers) of a transformed positive database in some extended language.     Moreover the computation of  minimal answers in the transformed database does not require the computation of cyclic trees.

For each cyclic tree  $\Cal T$ in $T$ introduce a new predicate $Q_{\Cal T}$, and let $\Cal L^* = \Cal L \cup \{Q_{\Cal T} \vert \Cal T$ is a cyclic tree in $T\}\cup \{{\tt FALSE}\}$.        For each $P\in \Cal L$, let $\phi(P)=\bigwedge\{Q_{\Cal T} \vert \Cal T$ is a cyclic tree for $P$ in  $T\}$.

For
each rule $C$: $
A_1 \wedge A_2 \wedge \ldots \wedge A_h \wedge \lnot A_{h+1}
\wedge \lnot A_{h+2} \wedge \dots \wedge \lnot A_{h+r} \to B_1
\vee B_2 \vee
\ldots \vee B_k
$, 
let $C^*$ be the rule 
$$A_1 \wedge A_2 \wedge \ldots \wedge A_h \wedge \phi(A_{h+1})
\wedge \phi(A_{h+2}) \wedge \dots \wedge \phi(A_{h+r}) \to B_1
\vee B_2 \vee
\ldots \vee B_k \vee {\tt FALSE}. 
$$
Let $T^* = \{C^* \mid C \in T\} \cup \{P\vee Q_{\Cal T} \vert P\in \Cal L, \Cal T$ is a cyclic tree for $P$  in $T
\}\cup T'$, where $T'=\{P\wedge \phi(P) \to {\tt FALSE} \,\vert\, P \in \Cal L\} \cup \{R\to
Q_{\Cal T} \mid \Cal T$ is a cyclic tree in $T, R\in \Cal
O(\Cal T) \cup \Cal N(\Cal T)\}.$

\p

If $T$ contains no rule $C$ for which antec$(C) = \emptyset$ then  $\emptyset$ is the unique disjunctive stable model of $T$ (in which case there are then no answers in $T$).    We may therefore assume that $T$ contains at least one rule $C_0$ for which antec($C_0)=\emptyset$ (whence antec($C_0)\cup \Cal N(C_0)=\emptyset$, cf., Section 1.1) and under these assumptions we note that 
$\Cal L^*-\Cal L$ is a minimal model of $T^*$.

Notice that $T^*$ is a positive database.  We show that (minimal) models of $T^*$ characterise disjunctive stable models of $T$.
Note also that $T'$ is definite (i.e., each rule's  head contains a single predicate), thus given $N\subseteq \Cal L^*$, let $cl(N)$ be the (unique) smallest model of $T'$ containing $N$.  Notice that if $M\subseteq \Cal L$, then $cl(M)\cap \Cal L = M$.

\p

\noindent {\bf B.1 Theorem.}

\item {\bf  (a)}  If $M$ is a disjunctive stable model of $T$, then $cl(M)\models T^* \wedge \lnot{\tt FALSE}$.

\item {\bf (b)}   Suppose that $N\subseteq \Cal L^*$ and $N\models T^*\wedge \lnot {\tt FALSE}$.  Let $M=N\cap \Cal L$, then $M$ is a disjunctive stable model of $T$ with $cl(M)\subseteq N$.

\p

\noindent {\bf Proof (a)}    If $P\in M$, then there is some cyclic tree $\Cal T$ for $P$ in $T$ such that $[\Cal O(\Cal T)\cup \Cal N(\Cal T)]\cap M =\emptyset$, whence $Q_{\Cal T}\not\in cl(M)$ (and in particular $\phi(P)$ is false in $cl(M)$).   Similarly if $P\in \Cal L - M$, and $\Cal T$ is a cyclic tree for $P$ in $T$, then by Theorem 2.4(c) we must have that $[\Cal O(\Cal T)\cup \Cal N(\Cal T)]\cap M \ne \emptyset$ (else $Pred(\Cal T) \subseteq M$), whence  $Q_{\Cal T}\in cl(M)$.  

It thus follows trivially that $cl(M)\models T^*\wedge \lnot {\tt FALSE}$.

\noindent {\bf (b)}  Suppose that $C\in T$ with $\text{antec}(C)\subseteq M$ and $M\cap \Cal N(C)=\emptyset$.  If $P\in \Cal N(C)$, then $P\not \in N$, whence $Q_{\Cal T}\in N$ for each cyclic tree $\Cal T$ for $P$ in $T$.   Since $N\models C^*$ it then follows that $N\cap \text{conseq}(C^*)=M\cap \text{conseq}(C)\ne\emptyset$.

Suppose that $P\in M$.  Since ${\tt FALSE}\not \in N$ there is a  cyclic tree $\Cal T$ for $P$ in $T$ such that $Q_{\Cal T}\not\in N$, whence $[\Cal O(\Cal T)\cup \Cal N(\Cal T)]\cap M = \emptyset$.   By Theorem 2.4(c), $Pred(\Cal T) \subseteq M$ and by Theorem 2.6,  $\overline{M}\cup (\Cal L - M)$ is cyclic and $M$ is a disjunctive stable model.  \bb

\p

We may then easily obtain a one-to-one mapping between disjunctive stable models of $T$ and minimal models of $T^*\wedge \lnot {\tt FALSE}$.

\p

\noindent {\bf B.2 Corollary.}  

\item {\bf (a)}    If  $M\subseteq \Cal L$, then $M$ is a disjunctive stable model of $T$ iff 
$cl(M)$ is a minimal model of $T^*\wedge \lnot {\tt FALSE}$.   

\item {\bf (b)}  Every minimal model $N$ of $T^*\wedge \lnot {\tt FALSE}$ has the form $N=cl(N\cap \Cal L)$, where $N\cap \Cal L$  is a disjunctive stable model of $T$.

\p

\noindent {\bf Proof (a).}  If $M$ is a  disjunctive stable model of $T$, then by the above theorem, 
$cl(M)\models T^*\wedge  \lnot {\tt FALSE}$.    Suppose that $N\subseteq cl(M)$ is a  model of $T^*$, then by Theorem B.1(b),  $N\cap \Cal L$ is a disjunctive stable model of $T$ with $N\cap \Cal L \subseteq cl(M)\cap \Cal L = M$.  By the minimality of $M$ we must have that $M=N\cap \Cal L$, and hence that $cl(M)= cl(N\cap \Cal L)\subseteq N$.

For the converse, suppose that $cl(M)$ is a minimal model of $T^*\wedge \lnot {\tt FALSE}$.  By part (b) of the above theorem,  $M = cl(M)\cap \Cal L$ is a disjunctive stable model of $T$.

\noindent {\bf (b).}   If $N$ is a minimal model of $T^*\wedge \lnot {\tt FALSE}$, then  by the above theorem, $N\cap \Cal L$ is a disjunctive stable model of $T$, with $cl(N\cap \Cal L) \subseteq N$ and $cl(N\cap \Cal L) \models T^*\wedge \lnot {\tt FALSE}$.  By the minimality of $N$ we must have that $N=cl(N\cap \Cal L)$.    \bb

\p

\noindent {\bf B.3 Corollary.} 

\item {\bf (a)}   
If $\Cal A \subseteq \Cal L$, then  $T\models \bigvee \Cal A$ iff $T^* \models  {\tt FALSE} \vee \bigvee \Cal A$.

\item {\bf (b)}   If $\Cal A\subseteq \Cal L$, then   $\Cal A$ is a minimal answer in $T$ iff $\{{\tt FALSE}\}\cup \Cal A$ is a minimal answer in $T^*$.

\item {\bf (c)} $T$ has a disjunctive stable model iff $T^*\not\models {\tt FALSE}$.

\item {\bf (d)}  If $P\in \Cal L$, then $P$ belongs to some disjunctive stable model of $T$ iff $T^*\not\models {\tt FALSE} \vee \phi(P)$.

\p

\noindent {\bf Proof (a).}  Suppose that $T\models \bigvee \Cal A$, and that $N\models T^*\wedge \lnot {\tt FALSE}$.   By Theorem B.1(b),  $N\cap \Cal L\cap \Cal A \ne \emptyset$.   Conversely suppose that $T^* \models  {\tt FALSE} \vee \bigvee \Cal A$, and $M$ is a disjunctive stable model of $T$, then by Theorem B.1(a), $\emptyset \ne \Cal A\cap cl(M) =\Cal A \cap \Cal L \cap cl(M) = \Cal A \cap M$.  

\noindent {\bf (b).}      Suppose that $\Cal A$ is a minimal answer in $T$.    By the remarks preceding Theorem B.1, $\Cal L^*-\Cal L$ is a model of $T^*$, whence  $T^*\not\models \bigvee \Cal A$.  Thus  if $\Cal B\subset \{{\tt FALSE}\}\cup \Cal A$ is a minimal answer in $T^*$, then $\Cal B = \{{\tt FALSE}\}\cup \Cal B'$, where $\Cal B'\subset \Cal A$.  But then by part (a), $T\models \bigvee \Cal B'$, thus contradicting the minimality of $\Cal A$.

For the converse, suppose that $\{{\tt FALSE}\}\cup \Cal A$ is a minimal answer in $T^*$.    Part (a) dictates that $T\models \bigvee \Cal A$.  If $\Cal B \subset \Cal A$ with $T\models \bigvee  \Cal B$, then $T^*\models  {\tt FALSE}\vee \Cal B$, thus contradicting the minimality of $\{{\tt FALSE}\}\cup \Cal A$.  

The proofs of (c) and (d) are trivial.   \bb

\p

We can thus compute minimal answers in $T$ via minimal answers in $T^*$.  The following theorem (analogous to Theorem 3.3) provides the basis for such a computation, and also indicates that having computed cyclic trees in $T$, we do not need to repeat the process in $T^*$.

\p

\noindent {\bf B.4 Theorem.}   If $\Cal A = \{A_i \vert i \le r\}\subseteq \Cal L$, then $\Cal A$ can be extended to a minimal answer in $T$ iff for each $i\le r$ we may find a cyclic tree $\Cal T_i$ for $A_i$ in $T$ and a strong cover $\Cal C_i\subseteq \Cal L^*$ of $\{A_j \vert j\le r, j\ne i\}\cup \{Q_{\Cal T_i}, {\tt FALSE}\}$ in $T^*$ such that $T^*\models {\tt FALSE} \vee\bigvee_{i\le r} A_i \vee \bigvee (\Cal L \cap \bigcap_{i\le r} \Cal C_i)$.

\p

\noindent {\bf Proof $(\to)$.}  Suppose that $\Cal A'$ is a minimal answer in $T$ containing $\Cal A$, and that for each $i\le r$,  $M_i$ is a disjunctive stable model of $T$ such that $M_i\cap \Cal A'=\{A_i\}$.   Let $\Cal T_i$ be a cyclic tree for $A_i$ in $T$ for which $Pred(\Cal T_i)\subseteq M_i\subseteq \Cal L - (\Cal O(\Cal T_i) \cup \Cal N(\Cal T_i))$.   

By Corollary B.2(a), $cl(M_i)$ is a minimal model of $T^*\wedge \lnot {\tt FALSE}$ (whence $\Cal C_i = \Cal L^* - cl(M_i)$ is a strong cover in $T^*$).   It is easy to check that $\Cal C_i\supseteq \{A_j \vert j\le r, j\ne i\}\cup \{Q_{\Cal T_i}, {\tt FALSE}\}$.

Finally if $A\in \Cal A'-\Cal A$, then for each $i\le r$ $A\not\in M_i$, whence $A\not\in cl(M_i)$, thus $\{{\tt FALSE}\} \cup \Cal A'\subseteq \{{\tt FALSE}\}\cup \{A_i \vert i \le r\}\cup \bigcup (\Cal L \cap \bigcap_{i\le r} \Cal C_i)$, and the result then follows from Corollary B.3(a).

\noindent $(\leftarrow)$.   Let $\Cal B \subseteq \{A_i \vert i \le r\}\cup \bigcup (\Cal L \cap \bigcap_{i\le r} \Cal C_i)$ be such that $\{{\tt FALSE}\} \cup \Cal B$ is a minimal answer in $T^*$ (whence $\Cal B$ is a minimal answer in $T$).   But now $\Cal C_i$ is a strong cover in $T^*$, thus $\Cal L^* - \Cal C_i\models T^*$, and  $\emptyset \ne (\Cal L - \Cal C_i)\cap  (\{{\tt FALSE}\} \cup \Cal B)\subseteq \{A_i\}$.  Thus $\{A_i \vert i \le r\}\subseteq \Cal B$.     \bb

\p

Note that if we were to assume a partitioning of 
$\Cal L$ into EXT($\Cal L) \cup \text{INT}(\Cal L)$ (as in Section 5) then the pre-processing described above would {\it not} be immune to changes in INT($T$).   To accomplish such immunity we would (as in Section 5) need to employ cyclic trees  
in INT($T) \cup \{E \vert  E \in \text{EXT}(\Cal L)\}$, and this would allow us to transform the database into an equivalent database in which rules contain
only {\it extensional}  negative subgoals.

\end